\documentclass[10pt,journal,compsoc]{IEEEtran}
\usepackage{cite}
\usepackage{amsmath,amssymb,amsfonts}
\usepackage{algorithmic}
\usepackage{graphicx}
\usepackage{textcomp}
\usepackage{xcolor}
\usepackage[caption=true, font=footnotesize]{subfig}
\def\BibTeX{{\rm B\kern-.05em{\sc i\kern-.025em b}\kern-.08em
    T\kern-.1667em\lower.7ex\hbox{E}\kern-.125emX}}
\usepackage[utf8]{inputenc}
\usepackage{booktabs}
\usepackage{multirow}
\usepackage{tabularx}
\usepackage{tabulary}
\usepackage{url}
\usepackage{soul}
\usepackage{verbatim}
\usepackage{booktabs}
\usepackage[np, autolanguage]{numprint}
\usepackage{nicefrac}

\usepackage{tikz}
\newcommand*\circled[1]{\tikz[baseline=(char.base)]{
            \node[shape=circle,draw,inner sep=1pt,font=\sffamily\footnotesize] (char) {\textbf{#1}};}}
\usepackage{pifont}

\usepackage{pdfcomment}

\usepackage{soul}
\usepackage{color}
\usepackage{pdfcomment}
\usepackage{todonotes}

\ifCLASSINFOpdf
\else
\fi

\begin{document}
\title{Fault Injection Analytics: A Novel Approach to Discover Failure Modes in Cloud-Computing Systems}

\author{Domenico Cotroneo,
        Luigi~De~Simone,
        Pietro~Liguori,
        and~Roberto~Natella}% <-this % stops a space

\markboth{}%
{}

\IEEEtitleabstractindextext{%
\begin{abstract}
Cloud computing systems fail in complex and unexpected ways due to unexpected combinations of events and interactions between hardware and software components. Fault injection is an effective means to bring out these failures in a controlled environment.
However, fault injection experiments produce massive amounts of data, and manually analyzing these data is inefficient and error-prone, as the analyst can miss severe failure modes that are yet unknown. 
This paper introduces a new paradigm (\emph{fault injection analytics}) that applies unsupervised machine learning on execution traces of the injected system, to ease the discovery and interpretation of failure modes. 
We evaluated the proposed approach in the context of fault injection experiments on the OpenStack cloud computing platform, where we show that the approach can accurately identify failure modes with a low computational cost.
\end{abstract}

}

% make the title area
\maketitle

\IEEEdisplaynontitleabstractindextext

\IEEEraisesectionheading{
\section{Introduction}
\label{sec:introduction}
}
\begin{figure*}[t]
\centering
\includegraphics[width=2\columnwidth]{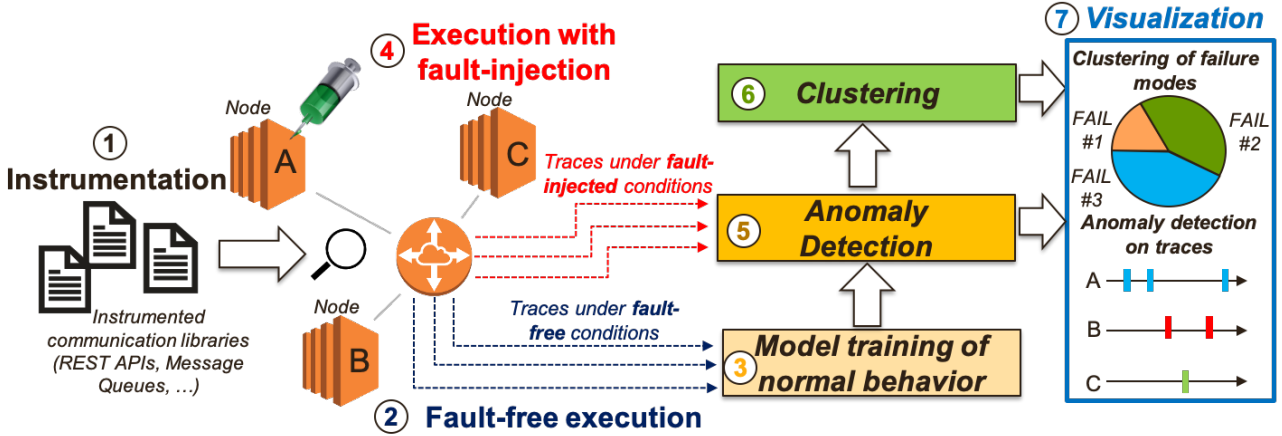}
\caption{Overview of the proposed approach.}
\label{fig:approach_overview}
\end{figure*}

Cloud computing has grown rapidly in recent years. It is well known that failures in these systems might have huge financial implications for the companies involved and their customers. Unfortunately, cloud-computing systems fail in complex and unexpected ways. For instance, recent outages reports showed that failures escape fault-tolerance mechanisms, due to unexpected combinations of events and of interactions among hardware and software components, which were not anticipated by the system designers \cite{garraghan2018emergent,hole2019software}. 
These issues become elusive due to the (large) scale of systems (with often thousands of nodes), their heterogeneity, and the high variability of workloads. In this context, understanding how the system can fail (i.e., its \emph{failure modes}) is a crucial activity to define proper recovery strategies.

Fault injection has been advocated as an effective means to analyze failures of distributed systems in a controlled environment, by forcing faults and exceptional conditions. 
In the current approaches, analysts write failure specifications before the experiments. Then, they look for occurrences of these failures within the experimental data \cite{wolter2012resilience}. For example, the most sophisticated approaches check formal specifications over events and outputs, by using finite state machines \cite{deligiannis2016uncovering}, temporal logic predicates \cite{arlat2002dependability}, relational logic \cite{gunawi2011fate}, and special-purpose languages \cite{reynolds2006pip}. 
Since these specifications are mostly based on prior knowledge and experience of system designers about failures, they are not meant for discovering new, unknown failure modes of a distributed system, which are missed by the failure specifications. Moreover, writing failure specifications is a time-consuming and cumbersome task, which makes fault injection less applicable in practice.

In this work, we introduce a new paradigm to data analysis for fault injection experiments, which we call \emph{fault injection analytics}. Our approach combines \emph{distributed tracing} to gather raw failure data, and \emph{unsupervised machine learning} to discover the failure modes of the injected system. 

The approach aims to make easier, for human analysts, the identification of the failure modes among large amounts of data produced by fault-injection experiments. When considering complex cloud systems, it is typical to perform a large number of experiments (e.g., several thousand), since these systems include tens of processes and nodes and millions of lines of source code in which faults can be injected. Moreover, for each experiment, the system generates high volumes of log files (up to hundreds of MBs) and long execution traces (e.g., thousands of events per trace). Thus, it is not feasible in practice for the analyst to analyze all of these data in a reasonable amount of time.

The paper brings the following contributions.

\vspace{2pt}
\textit{1)} A novel \textbf{anomaly detection algorithm} to find unusual events and interactions (i.e., symptoms of failures) that occurred in fault injection experiments. We designed the algorithm to be robust to noise in cloud systems, caused by \emph{non-determinism} of timing and order of events, and to be quickly trained only a small set of \emph{fault-free} executions of the distributed system, by using a \emph{variable-order Markov Model}. Anomaly detection can aid human analysts at scrutinizing more efficiently the events that occurred during an experiment, by discarding uninteresting events, e.g., unusual, yet benign orderings of events caused by non-determinism. 
    
\vspace{2pt}
\textit{2)} An unsupervised method to \textbf{classify the failure modes}, which combines clustering with the proposed anomaly detection algorithm in order to automatically identify the failure classes among large sets of fault injection experiments. 
This approach allows human analysts to find recurring failure patterns and to add new fault-tolerance mechanisms for them. It is sufficient for the analyst to only analyze one or a few experiments from the same class, thus making the analysis  more efficient.
    
\vspace{2pt}
\textit{3)} An \textbf{experimental case study} on the widespread \emph{OpenStack} cloud management platform \cite{OpenStackProducts, OpenStackUsers}. We targeted the three main sub-systems of OpenStack (Nova, Neutron, Cinder) with fault injection under several scenarios. We found that anomaly detection can pinpoint anomalous events with a high hit rate, and can halve the number of false alarms due to non-determinism. Moreover, the anomaly detection algorithm noticeably improves the accuracy of clustering at identifying failure modes.

In our previous work \cite{cotroneo2019failviz,cotroneo2019enhancing}, we implemented a basic toolkit for collecting and visualizing distributed traces, and we performed a preliminary analysis to assess the feasibility of anomaly detection. 
This paper extends previous work by presenting an extensive experimental evaluation, and by addressing the previously-unexplored problem of failure mode clustering. 

This work is at the intersection of fault injection, machine learning (ML), and cloud computing. Most of the studies in these fields applied machine learning during the operational phase of the lifecycle of cloud systems, in order to detect, predict, and diagnose failures in cloud infrastructures \cite{lin2018predicting,nedelkoski2019anomaly,mariani2018localizing,ye2018fault,dean2012ubl}. These studies applied fault injection to validate the effectiveness of novel anomaly detection techniques. Other studies adopted machine learning to generate functional and fault injection tests \cite{meinke2015learning}. 
Our approach is unique since we apply machine learning \emph{before} a system is put in operation, to ease the analysis and the interpretation of huge amounts of data produced by fault injection, thus, the name \emph{fault injection analytics}. 
Existing ML approaches could not be applied for this goal, since they are expected to be trained with relatively-large training datasets to achieve high accuracy. 
In the testing phase, such large datasets are not affordable, due to the characteristics of modern software development, with a high frequency of releases and limited time for testing. Developers typically run nightly automated tests, which need to complete within a few hours in order to meet release deadlines. Therefore, we designed our proposed approach to be applicable with a low amount of training traces.

In the following, Section~\ref{sec:method} presents the proposed approach; Sections~\ref{sec:evaluation} and \ref{sec:clustering} experimentally evaluate anomaly detection and the clustering of failure modes; Section~\ref{sec:related} discusses related work; Section~\ref{sec:conclusion} concludes the paper.

\section{Proposed Methodology}
\label{sec:method}
\begin{figure*}[t]
\centering
\includegraphics[width=2\columnwidth]{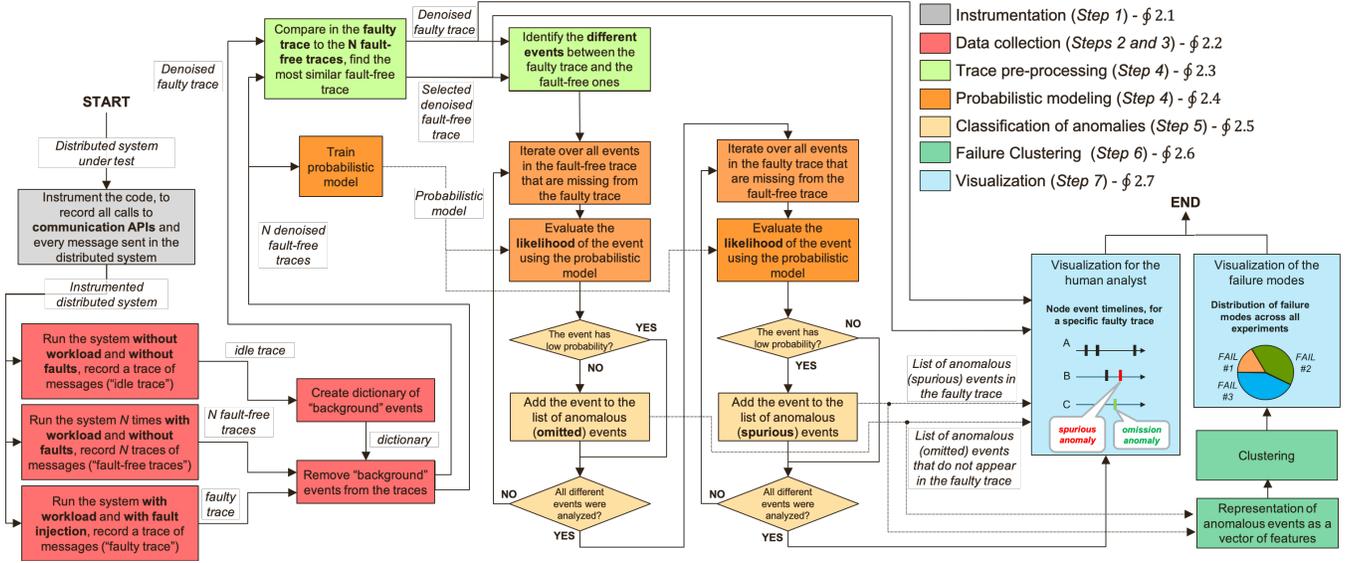}
\caption{Detailed workflow of the anomaly detection.}
\label{fig:approach_flowchart}
\end{figure*}

Our approach analyzes the cloud-computing system as a set of \emph{black-box} communicating components, without leveraging any a priori information about their internals (e.g., the approach is unaware of invariants and pre-/post-conditions in the system). Thus, we apply unsupervised machine learning on execution traces to identify failure patterns. 

The approach focuses on \emph{messages} exchanged in the distributed system during the fault injection experiments. 
In general, messages are the key observation point for debugging and verification of distributed systems, since they reflect well the activity of the system \cite{leesatapornwongsa2016taxdc}. For example, nodes perform work when they receive messages to provide a service to another node (e.g., through remote procedure calls), and reply with messages to provide the response and results; moreover, nodes use messages to asynchronously notify a new state to other nodes in the system. 
Our approach is plugged into the public communication interfaces, such as REST APIs and message queues, based on off-the-shelf protocols and libraries, and it collects raw traces of messages exchanged among the components.

An important design objective is to make the approach \emph{robust to non-determinism in distributed systems}, where the timing and the order of messages can unpredictably change (e.g., due to sporadic delays) regardless of the occurrence of failures. 
Thus, there is a need to discriminate between variations in the message order due to failures and by ``benign'' variations caused by non-determinism. To mitigate this uncertainty, we adopt a probabilistic model for anomaly detection that screens out the benign variations.

Another design objective is to use as few training samples as possible. Our approach trains a model by executing the system several times. However, since the execution time to run a cloud workload can be significantly long (e.g., in our experiments, a single run takes tens of minutes), it is mandatory to keep these runs at a minimum to make the approach affordable in practice.

\figurename{}~\ref{fig:approach_overview} shows an overview of the proposed approach. We first instrument communication APIs (step \circled{1}). Then, we exercise the system with a workload, and with no-fault injected (step \circled{2}). We record a trace of all messages exchanged among the components, and between the components and the clients. Since no fault is injected, such trace is denoted as \emph{fault-free trace}. We generate several fault-free traces, by running the workload several times. 
The fault-free traces are used as a training set to create a \emph{model of normal behavior} (step \circled{3}). We adopt a probabilistic model to account for the natural variability of the interactions (e.g., different ordering, type, etc.) in the training traces. 
%
%%%REVISION 2.2

We remark that having a representative experimental environment (i.e., matching the real-world operational environment, in terms of user workload, hardware, etc.) is a problem not limited to our approach, but it is a more general problem for fault injection \cite{basiri2016chaos}. 
Our goal is to facilitate the analysis of fault injection data, regardless of how well the data matches the operational environment (e.g., by architecting a realistic workload, by using a realistic configuration, etc.).

Once the model has been trained, the approach performs the fault injection experiments (step \circled{4}). We focus on injecting one fault per experiment, as injecting multiple faults concurrently is still an open research problem and not yet adopted in real projects, due to the high number of combinations among multiple faults.
This step produces \emph{fault-injected traces} (also \emph{faulty traces}), one per experiment. The fault-injected traces are then analyzed using the previously defined normal behavior model to identify anomalies (step \circled{5}). 
Since all the executions (i.e., fault-free and fault-injected ones) are performed under the same conditions (i.e., same software and hardware configuration, same workload, etc.), any deviation between a fault-injected trace and the probabilistic model is attributed to the injected fault and it is considered as an anomaly. The results of anomaly detection (i.e., the deviations between a fault-injected trace and the model) are the input of the clustering phase (step \circled{6}). This step aims to partition fault injection experiments in a number of groups such that experiments belonging to the same group exhibit the same anomalies (i.e., \emph{failure mode}).
Finally, the failure modes are visualized to the human analyst (step \circled{7}), by displaying the distribution of failure modes across all the experiments. Moreover, the user can focus on a specific experiment, by visualizing the anomalies of the execution over timelines. 

The anomaly detection algorithm constitutes the core of the proposed approach.  Figure~\ref{fig:approach_flowchart} shows a detailed flowchart of this algorithm. In the rest of this section, we discuss the phases of the workflow and present an example of fault injection analytics of a real system.

\subsection{Instrumentation}

The first step of the approach consists of instrumenting the system under test, to collect the messages exchanged by nodes during the experiments \cite{sigelman2010dapper}. 
To this purpose, our approach wraps the \emph{communication APIs} that are invoked by every component in the system. 

This instrumentation is a form of ``black-box tracing'', since it does not require any knowledge about the internals of the system under test, but it requires only basic information about the communication APIs being used. This approach is especially suitable when testers may not have a full and detailed understanding of the entire cloud platform.  
Moreover, this kind of distributed tracing is already familiar to developers for debugging, performance monitoring and optimization, root cause analysis, and service dependency analysis \cite{chow2014mystery,chen2002pinpoint}.

The information recorded by the instrumented APIs includes the time at which a communication API has been called and its duration; the component that invoked the API (\emph{message sender}) and the remote service that has been requested through the API call (\emph{service API}). Moreover, we record information about the response message (e.g., the status line and the message body in an HTTP response, the body of the message, etc.). 
We refer to the calls to communication APIs (i.e., the messages collected during the experiments) as \textbf{events}.
Thus, the execution of the system generates a \textbf{trace} of events that are ordered with respect to the timestamp given by the event collector.

Our anomaly detection technique is designed to be tolerant to the non-determinism in the ordering of the events (e.g., due to random messaging delays) by using a probabilistic technique, which is discussed in the Section \ref{subsec:probabilistic_modeling}.

\subsection{Data collection}

Once the system has been instrumented, it is executed with the workload, collecting traces without injecting any fault (\textbf{fault-free traces}). Such fault-free traces (also known as \emph{golden runs} or \emph{reference runs}) have been adopted for fault injection experiments in small systems (e.g., embedded ones), by using the traces as a reference to understand how the fault-injected system derailed from a proper execution \cite{leeke2009evaluating,lemos2012specification,natella2016assessing}. We generalize this approach to support more complex systems, such as cloud computing ones and use unsupervised machine learning to discover unknown failure modes. 
In the next steps, we will use fault-free traces to train a model of ``normal'' behavior of the distributed system, which we will use as a reference for analyzing failures. The model takes into account the variability of events across executions of the system (e.g., differences in the relative ordering of messages). Then, the system is executed again under fault injection, using the same workload of fault-free runs. For each experiment, we inject a different fault, and we collect a trace (\textbf{faulty trace}) of the events that are generated during the execution. Thus, we obtain several traces, one per experiment.

To recognize events that are generated by background and asynchronous activities, which are independent of the workload, we collect a third type of trace, namely (\textbf{idle trace}), which contains events occurring in the distributed system not caused by the workload or by the injected faults. Indeed, if these events are not removed from our analysis, they might be erroneously identified as (false) anomalies. 
Examples of such events are garbage collection, resource monitoring, updating database indexes, etc., and they can be triggered at arbitrary times. Another example in the OpenStack cloud computing platform is the events generated by the invocation of the method \textit{sync\_instance\_info} of the \textit{Nova Scheduler} component: this method is periodically called by compute nodes to notify the UUIDs of instances on the hosts, and it is not related to the workload.

To identify these events, we perform a separate execution of the cloud system, by leaving it in idle state (i.e., no workload is applied) for several minutes before and after a fault-free execution. We record into the idle trace any background message collected during these periods. Then, we remove such background events from both the fault-free and faulty traces.

\subsection{Trace pre-processing}
\label{subsec:trace_comparison}

Each event in the system is described by the couple of $<$\emph{message sender}, \emph{service API}$>$. In our context, the \emph{service API} represents the name of the invoked method  (e.g., \emph{create volume}), whereas the \emph{message sender} is the name of the sub-system invoking the method (e.g., \emph{Cinder}). 
The proposed approach represents the events within a trace with unique identifiers (i.e., \emph{symbols}), so that two events of the same type are identified by the same symbol.  Besides the specific event, we also consider the response status in the assignment of the symbols. For example, if the event is an HTTP message, we differentiate among invocations of the same GET method with different status codes (e.g., \numprint{200} for success, and \numprint{404} for failure).
Events in a trace are ordered by their time of collection, and then converted into \emph{sequences of symbols}: each symbol represents a specified couple $<$\emph{message sender}, \emph{service API}$>$, and the response status.

Once all execution traces have been converted into sequences, before resorting to anomaly detection, we perform preliminary filtering of events that do not represent anomalies. We identify events that do not exhibit any difference between the fault-injected and the fault-free executions, i.e., events that occur regardless of the injected fault. Since these events are not related to the failure modes, they can be discarded from the analysis. To identify these events, we look for overlapping symbols (i.e., same type, same order) between the faulty sequences and the fault-free ones. 

The approach identifies overlapping symbols between the sequences, by computing the \emph{longest common sub-sequence} (LCS) of the sequences \cite{bergroth2000survey}: 
the LCS is a subset of symbols that are present in both sequences in the same order, and that can be obtained by removing (a minimal number of) symbols from the original sequences. This kind of problem is recurrent in computer science, such as in bioinformatics and source code versioning (e.g., in the \emph{diff} Unix tool), and can be solved with efficient algorithms \cite{hunt1976algorithm,myers1986nd}.
The approach identifies a \emph{selected fault-free trace} that is \emph{most similar} to the fault-injected trace, i.e., the one with most overlapping symbols, by computing the normalized length of the LCS \((nLCS)\) between the faulty trace and the fault-free ones, where $nLCS(x,y)=\frac{|LCS(x,y)|}{\sqrt{l_x \cdot l_y}}\label{eq:normalized_lenght}$,
and where $l_x$ and $l_y$ are the lengths of the individual strings $x$ and $y$. Then, it generates a list of differences (i.e., non-common events) between the selected fault-free trace and the faulty trace. These non-common events are further analyzed with a probabilistic model, to tell which ones are indeed anomalies.

\subsection{Probabilistic modeling}
\label{subsec:probabilistic_modeling}

The analysis performed with LCS still does not to suffice to identify failure-related events, since the differences in the faulty trace can be either actual symptoms of a failure (i.e., real anomalies, caused by the injected fault); or non-anomalous events (i.e., events that may, or may not occur in fault-free conditions, or may occur in a different order, due to non-deterministic behavior). 
The latter type of events may lead to \emph{false alarms}, which may divert the attention of the human analyst. 
To overcome inaccuracies, we use a probabilistic model, in cascade after the trace analysis with LCS, to evaluate whether a non-common event is indeed an anomaly.

In particular, our approach uses a \emph{Markov model} to estimate the probability of an event. Markov modeling is a popular approach for the probabilistic analysis of sequences of symbols (e.g., to predict the probability of a future symbol), such as in bioinformatics \cite{stanke2003gene}, data compression \cite{rissanen1983universal}, and text and speech recognition \cite{rabiner1989tutorial}. 
Markov models do not require massive datasets to be trained, which is instead the case for other anomaly detection techniques like neural networks. The size of the training set is an important concern in our context, as developers have a limited time budget to spend for fault injection testing \cite{arlat2002dependability}. Since executions can take several hours in commercial-grade systems, we need to train the model with a minimal number of fault-free executions.

Among Markov models, \emph{Hidden Markov Models} (HMMs) are a powerful and very popular technique among researchers in dependable computing, such as for anomaly detection and fault diagnosis purposes in critical infrastructures \cite{basile2006approach,zonouz2012scpse,daidone2006hidden,carnevali2019learning}. HMMs separate \emph{observations} (e.g., events) from the (hidden) \emph{states} of the underlying stochastic process that generates the observations, since in many systems the current state is unknown for an external observer, and must be indirectly inferred from events \cite{rabiner1989tutorial}. 
%Therefore, HMMs can be used to estimate the probability of a sequence of observed events; and the probability of a sequence of states, given a sequence of observed events. 
However, we found that HMMs are not suitable for our anomaly detection problem. The main issue with HMMs is the high flexibility of the model, in terms of the high number of parameters that need to be tuned in the training phase (e.g., the number and probabilities of the hidden states). 
During the training phase, we cannot rely on a human analyst to annotate the events with the corresponding hidden state of the system, as it would be exceedingly time-consuming and error-prone for complex distributed systems with many unknown states. Instead, training HMMs with unannotated traces significantly increases the required size of the training set (e.g., up to thousands of traces using the EM algorithm) \cite{carnevali2019learning}. 
Another issue is the \emph{zero frequency problem}, that is, modeling the probability of events with no occurrences in the training set, which is often the case in anomaly detection \cite{witten1991zero}.

%.....which contains a set of hidden states \(X_i\) that produce a  symbol \(Y_i\)  at each time.  
%In the HMM algorithm, the output of the model only depends on the current state of the system so the previously observed symbols would not have any effects on the output.
Therefore, we opt for a non-hidden Markov model where the states are a direct representation of the observed events. However, a simple Markov chain still does not suffice for our purposes, since the probability of the next state (i.e., the next event of the sequence) would only depend on the current state (i.e., the \emph{memoryless} property). In general, this is not the case for event sequences that can be generated by a distributed system; in practice, the probability of an event is highly correlated with the history of the previous events. For example, in the OpenStack platform, the occurrence of an event representing a ``volume attach'' operation must be preceded by a sequence of several preliminary operations on the volume and on the instance to be attached (e.g., an instance must have been created and initialized).

Ultimately, we opted for \emph{higher-order} Markov models, where the probability of events takes into account the history of the previous states of a sequence. In particular, since we do not have a fixed cardinality for the conditioning set of events in history, we adopt \emph{Variable-order Markov Models} (VMMs). VMMs estimate the probability that a symbol \(\sigma\) can appear after a sequence $s$ (named \emph{context}), by counting the joint occurrences of \(\sigma\) and $s$ in the training sequence to build the predictor \(\hat{P}\), for variable cardinalities of $s$ \cite{begleiter2004prediction}.

%We used the VMM algorithm rather than \textit{neural networks} since the latter can require a high number of training sequences, which can be costly to get (it takes several tens of minutes to execute an intensive workload). The VMM algorithm can provide good accuracy even with few training traces, as shown in \S{}~\ref{subsec:accuracy}.

%Ultimately, we opted for \emph{higher-order} Markov model, where the probability of events takes into account the history of the previous states of a sequence. In particular, since we do not have a fixed cardinality for the conditioning set of events in history, we adopt \emph{Variable-order Markov Models} (VMMs). VMMs estimate the probability that a symbol \(\sigma\) can appear after a sequence $s$ (named \emph{context}), where the context length \(|s|\) vary in response to the available statistic in the training sequence \cite{begleiter2004prediction}. These counts provide the basis for generating the predictor \(\hat{P}\).

%In contrast to the Markov chain models, where each random variable in a sequence with a Markov property depends on a fixed number of random variables, in VMMs this number of conditioning random variables may vary depending on the length of the context \cite{begleiter2004prediction}.

In this work, we use the notation defined by Begleiter et al. \cite{begleiter2004prediction}.
Let \(\Sigma\) be a finite alphabet. A learner is given a training sequence \(x_1^n = x_1 x_2 ... x_n\), where \(x_i \in \Sigma\) and \(x_i x_{i+1}\) is the concatenation of \(x_i\) and \(x_{i+1}\). Based on \(x_1^n\), the goal is to learn a model \(\hat{P}\) that provides a probability assignment for any future outcome given some past.
Specifically, for any context \(s \in \Sigma\) and symbol \(\sigma \in \Sigma\), the learner should generate a conditional probability \(\hat{P}(\sigma|s)\). 
The accuracy of the predictor \(\hat{P}(\cdot|\cdot)\) is typically measured by its average log-loss \(l(\hat{P},x_1^T)\) with respect to a test sequence \(x_1^T = x_1 ... x_T\):

\begin{equation}
    \ell(\hat{P},x_1^T)= -\frac{1}{T} \sum_{i=1}^{T} \log \hat{P}(x_i | x_1...x_{i-1})
\end{equation}

%where the logarithm is taken to base $2$. 

%The average log-loss is directly related to the following likelihood:
%\begin{equation}
%    \hat{P}(x_1^T)=\prod_{i=1}^{T}\hat{P}(x_i | x_1...x_{i-1})
%\end{equation}
% Minimizing the average log-loss is completely equivalent to maximizing a probability assignment for the entire test sequence.

%There are many VMM prediction algorithms. Any lossless compression algorithm can be used for prediction since there is a relation between the prediction of discrete sequences and lossless compression algorithms \cite{begleiter2004prediction}. 

There exist many algorithms in the scientific literature for training and applying VMMs \cite{begleiter2004prediction}. In particular, one important aspect that characterizes VMM algorithms is how they handle the zero-frequency problem (i.e., sequences with zero occurrences in the training set). If the probability is estimated by simply counting the number of occurrences, the unobserved events would get a zero probability, with an infinite log-loss. This problem is especially relevant in the case of long sequences with a rich alphabet, where the training set is ``sparse'' and only covers a tiny part of the multi-dimensional space of the sequences. The sequence of events generated by a distributed system also falls in this condition.

Our approach uses the \emph{Prediction by Partial Matching, Method C} (PPM-C) lossless compression algorithm \cite{cleary1997unbounded}, which is a variant of the original PPM algorithm published in 1984 by Cleary and Witten \cite{cleary1984data} that includes a set of improvements proposed by Moffat \cite{moffat1990implementing}. 
% "finite-context"
PPM is a statistical modeling technique that builds a predictor by combining several fixed-order context models \cite{cleary1997unbounded}, with different values of the order $k$, ranging from zero to an upper bound $D$ (i.e., the maximal order of the Markov model) \cite{mavadati2014comparing}.

%All PPM variants handle the zero-frequency problem by using two mechanisms, called \emph{escape} and \emph{exclusion}. For each context $s$ of length \(k \leq D\), the algorithm allocates a uniform probability mass \(\hat{P_k}(escape|s)\) for all symbols that did not appear after the context $s$ in the training sequence. The remaining mass \(1-\hat{P_k}(escape|s)\) is distributed among all other symbols that have non-zero counts for this context.

All PPM variants manage the zero frequency problem by using two mechanisms, called \emph{escape} and \emph{exclusion}. For each context $s$ of length \(k \leq D\), the algorithm allocates a uniform probability mass \(\hat{P_k}(escape|s)\) (which varies across PPM variants) for all symbols that did not appear after the context $s$ in the training sequence. The remaining mass \(1-\hat{P_k}(escape|s)\) is distributed among all other symbols that have non-zero counts for this context. Using the escape mechanism, the conditional probability is given by \cite{begleiter2004prediction}:

%\begin{equation}
%    \hat{P}(\sigma| s^{n}_{n-D+1})=
%\end{equation}

\begin{align}
  \hat{P}(\sigma| s^{n}_{n-D+1})  & = \begin{cases} 
        \hat{P}_D(\sigma| s^{n}_{n-D+1}), & ^\dagger{} \\
        \hat{P}_D(escape| s^{n}_{n-D+1})\cdot\hat{P}(\sigma| s^{n}_{n-D+2}), & ^\ddagger{}\\
    \end{cases}
\end{align}

{
\small
\noindent 
\hfill$^\dagger$ $\mbox{if } s^{n}_{n-D+1}\sigma$ occurred in the training sequence \hfill$^\ddagger$ otherwise
}

\vspace{3pt}

%...... = 1/|\Sigma|
%......inversely proportional to the size of the alphabet

%        & \mbox{if } s^{n}_{n-D+1}\sigma
%        & \mbox{appeared in the}\\
%        & \mbox{training sequence;}&\\ 

\noindent
where $\hat{P}_D(\cdot|\cdot)$ is a conditional probability with fixed-order $D$, which can be calibrated according to frequency counts from the observed sequences in the training set.

The exclusion mechanism is used to tune the probability estimates. This probability is inversely proportional to the size of the alphabet (for example, the probability of the escape is $1/|\Sigma|$ in the case of an empty context $s = \epsilon$), but the PPM-C introduces a correction. If a symbol \(\sigma\) appears after the context $s$ of length \(k \leq D\), it is redundant to consider \(\sigma\) as part of the alphabet when computing \(\hat{P_k}(\cdot|s')\), for all $s'$ suffix of $s$. Therefore, the estimates \(\hat{P_k}(\cdot|s')\) are corrected by considering a smaller alphabet of observations \cite{begleiter2004prediction}. For more  information on PPM and the \textit{Method C} variant, we refer the reader to the work by Begleiter et al. \cite{begleiter2004prediction}.

%In our context (i.e., analyzing execution traces from distributed systems), we set the maximum order $D$ of the VMMs by measuring the number of events that can be triggered by an individual request from a client of the distributed system. In response to a client request, the distributed system generates a sequence of messages among its internal components, until it reaches again a quiescent state, or it returns a reply to the client. Since an event is most likely influenced by the surrounding events in the context of the same client request, we set $D$ to the maximum number of events triggered by a client request. This choice is a conservative approach since this number (e.g., several tens of events in our case study) tends to be much higher than the context length chosen in previous studies on VMMs; for example, Cleary et al. \cite{cleary1997unbounded} showed that PPM achieves the best compression with just a maximum context length of $D=5$, while a further increase of $D$ does not increase the accuracy, but only increases the computational time of the algorithm.

%Regarding the choice of the maximum order $D$, Cleary et al. \cite{cleary1997unbounded} showed that PPM achieves the best compression with just a maximum context length of $D=5$, while a further increase of $D$ does not increase the accuracy, but only increases the computational time of the algorithm.

We set the maximum order $D$ of the VMMs to $5$. Indeed, it has been found that PPM achieves the best compression for this choice and that its accuracy saturates when the context is increased beyond this value \cite{cleary1997unbounded}. %\hl{While longer contexts do provide more specific predictions, they also stand a much greater chance of not giving rise to any prediction at all.} 

%Cleary et al. in \cite{cleary1997unbounded} showed that the best compression for PPM is achieved when a maximum context length of five is chosen and that it deteriorates slightly when the context is increased beyond this. The reason is that, while longer contexts do provide more specific predictions, they also stand a much greater chance of not giving rise to any prediction at all. This causes the escape mechanism to be used more frequently to reduce the context length down to the point where predictions start to appear. And each escape operation carries a small penalty in coding efficiency.

\subsection{Classification of anomalies}
\label{subsec:classification}

The ultimate result of anomaly detection is to classify the events into:

\begin{itemize}
    \item\textbf{Common events}: Events that occurred both in the faulty trace and in at least one of the fault-free traces, with the same type and order.
    
    \item \textbf{Anomalous events}: Differences between the faulty trace and the fault-free traces. They are further classified into:
    \begin{itemize}
        \item \textbf{Spurious events}: Events that would normally not occur under fault-free conditions.
        \item \textbf{Missing events}: Events that occur in fault-free conditions, but do not occur under fault injection.
    \end{itemize}
    
\end{itemize}

As discussed in \S{}~\ref{subsec:trace_comparison}, we first use the LCS algorithm to identify common events of a faulty trace, by comparing it to a \emph{selected fault-free trace} (i.e., one of the fault-free traces in the training set, with the highest similarity to the faulty trace). Then, we further analyze the \emph{LCS differences} (i.e., non-common events according to the LCS) using the VMM model (\S{}~\ref{subsec:probabilistic_modeling}). 
We train the VMM with a set of $n-1$ fault-free traces, by using all the fault-free traces except the \emph{selected fault-free trace}. Then, we apply the VMM to compute the probabilities of LCS differences, to determine whether they are indeed anomalous, as follows:

\vspace{3pt}
\noindent
$\rhd$ \textbf{Analysis of LCS differences that only appear in the fault-injected trace.} 
The fault-injected trace takes the role of the \emph{test sequence} for the VMM. We focus on symbols of the test sequence that were highlighted as differences in the previous LCS analysis. The goal is to confirm whether these symbols are actually unlikely events, not only with respect to the selected fault-free trace (i.e., the one used for determining the LCS) but also according to the whole set of fault-free traces in the training set. For each event not included in the LCS, we compute the probability of the event according to the VMM. If the probability is lower than a threshold \(\epsilon_{\textrm{SPURIOUS}}\), then the symbol has a low likelihood to appear in that position of the sequence; thus, the VMM confirms that the symbol represents a \textbf{spurious} anomalous event. Otherwise, the event is considered non-anomalous.

\vspace{3pt}
\noindent
$\rhd$ \textbf{Analysis of LCS differences that only appear in the selected fault-free trace.} 
The \emph{selected fault-free trace} takes the role of the \emph{test sequence} for the VMM. As for the previous step, we focus on symbols of the test sequence that were highlighted as differences in the previous LCS analysis. In this case, we consider the events that only appear in the selected fault-free trace: therefore, from the point of view of the fault-injected trace, these events represent \emph{omissions}. This step confirms whether these events are likely, and thus their omission should be considered an anomaly. The approach applies the VMM to the events that only appear in the fault-free trace, by computing the probabilities of such events according to the remaining fault-free traces in the dataset. If the probability of the event is higher than a threshold \(\epsilon_{\textrm{MISSING}}\), then there is a high likelihood for the symbol to be in that position of the sequence. Therefore, the fact that the event is missing in the fault-injected trace should be considered an anomaly, and thus it is marked as a \textbf{missing} anomalous event. Otherwise, if the probability of the event is less than the threshold, then the lack of such an event from the fault-injected trace is considered non-anomalous.

\vspace{3pt}

We remark that even if the two steps perform similar comparisons, the results obtained by them are different and complementary. If the fault-injected trace contains an anomalous event with a \emph{low probability value} according to the VMM, then it is confirmed as spurious. Similarly, if the fault-injected trace does not contain an event with a \emph{high probability value} in the selected fault-free trace, then the event is confirmed to be an omission. 
A practical approach is to select conservative thresholds (e.g., \(\epsilon_{\textrm{SPURIOUS}} = 20\%\) and \(\epsilon_{\textrm{MISSING}} = 80\%\)), so that the VMM can filter out most of the LCS differences that are not actually spurious/missing events; and to leave to the human analyst the decision about the uncertain events. Therefore, the sensitivity of the probabilistic model is an important factor that makes it applicable in practice. We further analyze it in our experiments.

\subsection{Failure clustering}
\label{subsec:clustering_methodology}

The last step of our approach is to perform clustering to group the experiments into classes, where each class represents a distinct failure mode of the system under test. 
In general, clustering algorithms reveal hidden structures in a given data set, by grouping ``similar'' data objects together while keeping ``dissimilar'' data objects in separated groups \cite{xu2005survey}. In our context, the clustering of the experiments helps the human analyst in the identification of the failure modes and in analyzing the large amount of data of the fault-injection campaigns (hundred of MB of logs, thousand of traces and experiments, etc.).

To apply the clustering, the approach represents each fault-injection experiment with a vector of features. The number of features is twice the number $d$ of unique events (i.e., the symbols in the dictionary of events) that were traced during the experiments. 
Given that anomalies can be classified as spurious or missing, we include in the vector two features for each symbol: the number of times that the symbol occurred as a spurious anomaly (the first $d$ features), and the number of times that the symbol occurred as a missing anomaly (the last $d$ features). %Thus, the vectors include in total $2d$ features (i.e., twice the size of the dictionary). 
For example, let us suppose that the dictionary consists of three different symbols, $A, B, C$ (i.e., a dictionary with three unique events). Let be $x_i = [1,1,0,0,2,3]$ the vector associated to the faulty trace collected during the $i^{th}$ experiment. These features can be interpreted as follows:
\begin{itemize}
    \item Anomaly detection identified two spurious events, one for the symbol $A$ and one for the symbol $B$.
    \item Anomaly detection identified five missing events, two for the symbols $B$ and three for the symbol $C$.
\end{itemize}

This representation holds concise information about the anomalies of the experiments. 
Spurious events are indicators of wrong interactions that happened in the distributed system during the experiment while missing events point out actions that were not performed. 
We apply a clustering algorithm on these vectors, to group the experiments that exhibit similar anomalies.
Thus, clusters describe distinct failure modes exhibited by the system. Our approach is not bound to a specific clustering algorithm; we rely on the anomaly detection algorithm to detect the symptoms of the failures with high accuracy, in order to favor the quality of the failure clusters.

\begin{figure*}[!t]
\centering
    \subfloat[Distribution of failure modes.\label{fig:example_dashboard_distribution}]{%
        \includegraphics[width=0.6\columnwidth]{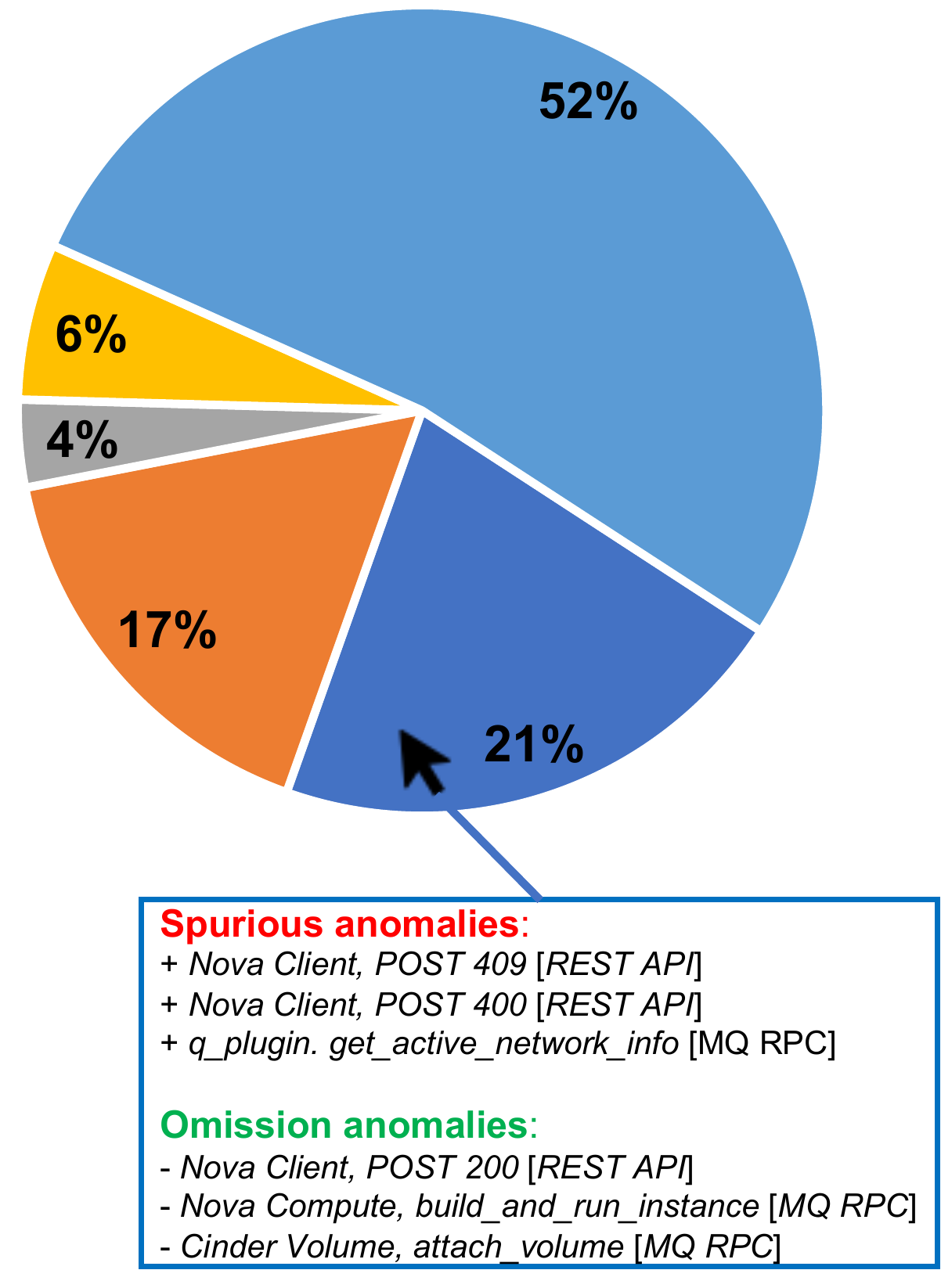}
    }
    \qquad
    \subfloat[Anomalous events in a specific fault injection experiment.\label{fig:example_dashboard_events}]{%
        \includegraphics[width=1\columnwidth]{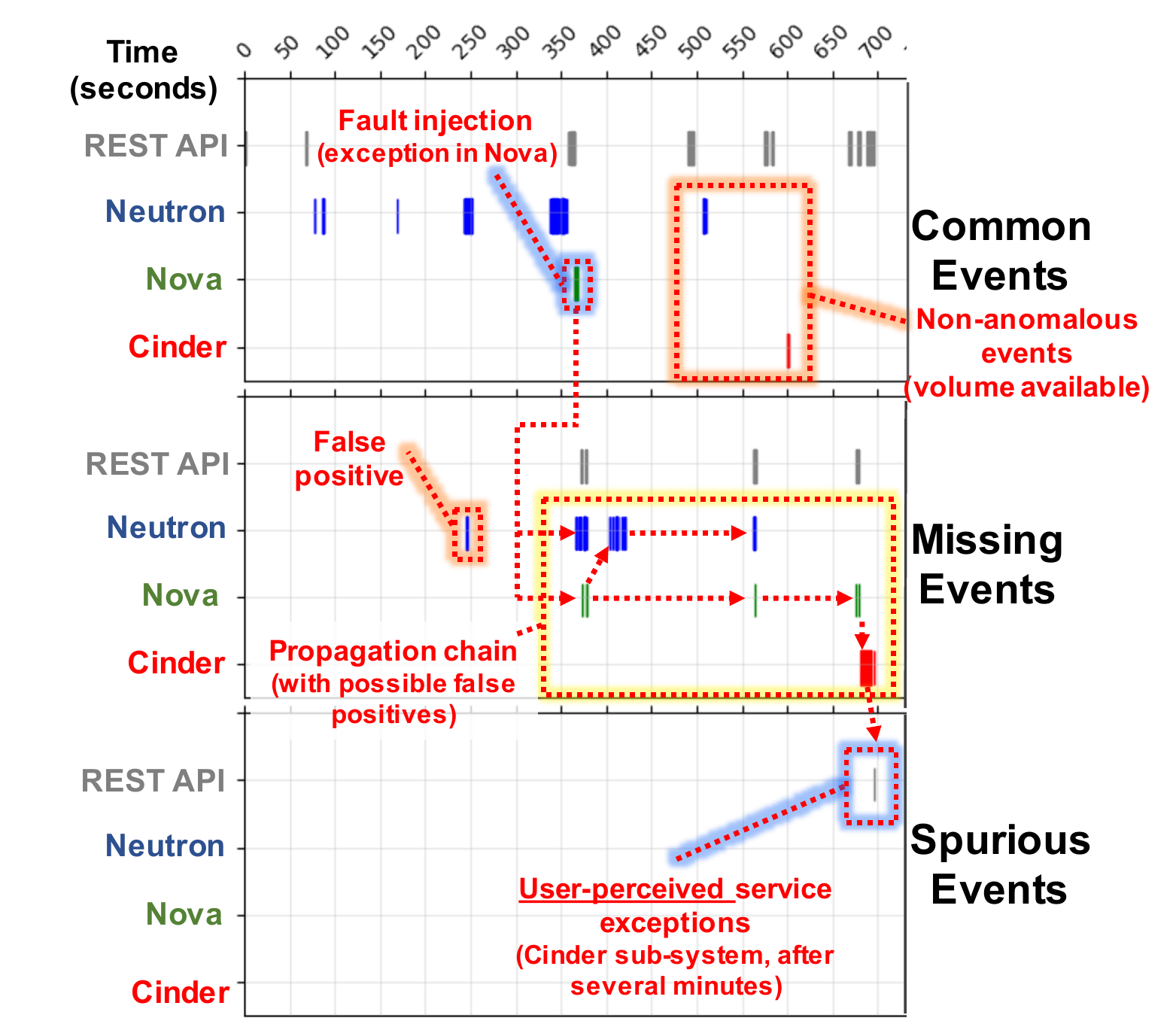}
    }

\caption{Example of fault injection data analysis.}
\label{fig:example_dashboard}
\end{figure*}

\subsection{Visualization}
\label{sec:problem}

Visualizing the execution of distributed systems is a key step to enable designers to debug failures, yet summarizing information in an effective way is an open research problem \cite{beschastnikh2020visualizing,beschastnikh2016debugging,barham2003magpie,reynolds2006pip}.
Therefore, we designed a dashboard to leverage unsupervised machine learning to obtain summarized information about failure modes, in order to present them in a simplified way. The dashboard does not require the user to manually configure the failure modes, thus supporting the analysis and discovery of unknown failure modes.

%Once the fault injection experiments have been executed and analyzed through anomaly detection and clustering, the results are presented to the user in a visual form, using a ``turn-key'' dashboard. Under the hood, we apply unsupervised machine learning on the experimental traces, without requiring the user to configure the failure modes. To illustrate how the proposed approach can be used in practice, we show an example of the presentation of results from fault injection analytics. 

Besides providing basic statistics about the experiments (e.g., number, duration), the first feedback for the user is the \emph{distribution of failure modes across the fault injection experiments}  (\figurename{}~\ref{fig:example_dashboard_distribution}). Both the categories (i.e., the failure modes) and their sizes (i.e., the number of experiments) are automatically generated through unsupervised machine learning. 
In the example of \figurename{}~\ref{fig:example_dashboard_distribution}, based on fault injections on the OpenStack platform, every failure mode is labeled with a summary of the spurious and omission anomalies occurred in that failure mode. 
%If the user selects one of the failure modes in the plot, the dashboard shows the list of experiments that exhibited the failure mode.
%From this list, the user can further select one of the experiments to get more detailed information. 
The dashboard groups the experiments into few classes (one per failure mode) to simplify the analysis of failure modes. The user can quickly get a better understanding of each failure mode, by only looking at one or a few experiments for that class.

%Otherwise, if the experiments were not automatically classified into failure modes, the user would need to analyze the whole set of unclassified experiments to obtain a similar interpretation of the results.

The dashboard also supports the user at inspecting \emph{anomalous events that occurred within individual experiments}. When the user selects an experiment, the dashboard displays the timespans of RPCs (e.g., message queues) and REST API calls. Timespans are divided with respect to the origin of the messages, such as the Nova, Neutron, and Cinder sub-systems and external clients in the case of OpenStack. 
The dashboard divides interactions among three groups, as defined in \S{}~\ref{subsec:classification}: \emph{common}, \emph{missing}, and \emph{spurious} events. %Common interactions are the ones that also happen during a normal (i.e., fault-free) execution of the system. Missing interactions are the ones that happen during normal execution but did not happen during the fault injection experiment. Vice versa, spurious interactions are the ones that only happened during the fault injection experiment.  
In the example shown in \figurename{}~\ref{fig:example_dashboard_events}, the spurious events are exceptions raised by two REST API calls. The missing events are internal calls to initialize a new VM instance and to attach virtual resources to it. Due to the injected fault in the Nova sub-system, it did not complete the initialization of the instance, leaving it in an inactive state, and propagated the problem to Neutron and Cinder. 
The visualization supports analysts at reasoning about how to best handle faults, e.g., when in the flow of interactions, and whether to manage it in Nova, Neutron, and/or Cinder. 

%Due to the injected fault in the Nova sub-system, it did not complete the initialization of the instance, leaving it in an inactive state. Both Nova and Neutron (the network manager) did not raise any API exception, but the failure only became apparent to the client later on, when invoking the API of the Cinder sub-system. In particular, after five minutes, the client experienced a service exception when calling the API of the Cinder storage manager, since it could not attach the instance to a virtual volume. The problem propagated across sub-systems, from Nova to Neutron and Cinder. 

%More information about this example is available on our technical report \cite{AAAAAA}. 
%More information about this example is available in our previous work \cite{cotroneo2019failviz}.

\section{Evaluation of Anomaly Detection}
\label{sec:evaluation}
We evaluate our anomaly detection algorithm with experiments on the OpenStack cloud management platform, which is a relevant case of a large and complex distributed system.

\subsection{Experimental setup}
\label{subsec:experimental_setup}
We injected faults into the three most important sub-systems of OpenStack \cite{denton2015learning,solberg2017openstack}: (i) \emph{Nova}, which provides services for provisioning instances (i.e., VMs) and handling their life cycle; (ii) \emph{Neutron}, which provides services for provisioning virtual networks, including resources such as floating IPs, network interfaces, subnets, and routers; and (iii) \emph{Cinder}, which provides services for managing block storage resources.
Each of these three sub-systems represents by itself a complex system, and they are developed as independent projects by distinct and dedicated teams \cite{pikeMetric1,pikeMetric2}.
We targeted OpenStack version 3.12.1 (release \emph{Pike}), deployed on Intel Xeon servers (E5-2630L v3 @ 1.80GHz) with 16 GB RAM, 150 GB of disk storage, and Linux CentOS v7.0, connected through a Gigabit Ethernet LAN.

In our tests, we injected faults during the interactions among OpenStack components. We targeted the internal APIs used by OpenStack components for managing instances, volumes, networks, and other resources. The injected faults represent exceptional cases, such as a resource that is not found or unavailable, a processing delay when retrieving a resource, or an incorrect value caused by the user, the configuration, or a bug inside OpenStack. In particular, we focus on the following types of faults:

\begin{itemize}

    \item \textbf{Throw exception}: An exception is raised on a method call, according to pre-defined, per-API list of exceptions.
    
    \item \textbf{Wrong return value}: A method returns an incorrect value. The wrong return value is obtained by corrupting the targeted object, depending on the data type (e.g., by replacing an object reference with a null reference, or by replacing an integer value with a negative one).
    
    \item \textbf{Wrong parameter value}: A method is called with an incorrect input parameter. Input parameters are corrupted according to the data type, as for the previous point.

    \item \textbf{Delay}: A method is blocked for a long time before returning a result to the caller. This fault can trigger timeout mechanisms inside OpenStack, and cause stalls.
    
\end{itemize}

We performed three distinct fault injection campaigns, in which we applied three different workloads:

\begin{itemize}
    
    \item \textbf{New deployment} (DEPL): This workload configures a new virtual infrastructure from scratch, by stimulating all of the target sub-systems (i.e., Nova, Cinder, and Neutron) in a balanced way. 
    This workload creates VM instances, along with key pairs and a security group; creates and attaches volumes to an existing instance; creates a virtual network and a subnet, with a virtual router; assigns a floating IP to connect the instances to the virtual network; reboots the instances, and then deletes them.
        
    \item \textbf{Network management} (NET): This workload includes network management operations, to stress more the Neutron sub-system and virtual networking. The workload initially creates a network and a VM and generates network traffic via the public network. After that, it creates a new network with no gateway, brings up a new network interface within the instance, and generates traffic to check whether the interface is reachable. Finally, it performs a router rescheduling, by removing and adding a router resource.
    
    \item \textbf{Storage management} (STO): This workload performs storage management operations on instances and volumes, to stress more the Nova and Cinder sub-systems. In particular, the workload creates a new volume from an image, boots an instance, then rebuilds the instance with a new image (e.g., as it would happen for an update of the image). Finally, it performs a cleanup of the resources. 
    
\end{itemize}

All the workloads invoke the OpenStack APIs provided by the Nova, Cinder, and Neutron sub-systems. We designed the workloads to cover several sub-systems of OpenStack and several types of virtual resources, similar to integration test cases from the OpenStack project \cite{openstack_tempest}, to point out potential failure propagation effects across sub-systems.

During the execution of the workload, any exception generated by API calls (\emph{API Errors}) is recorded. In-between calls to service APIs, the workload also performs \emph{assertion checks} on the status of the virtual resources, to point out failures of the cloud management system. 
These checks assess the connectivity of the instances through SSH and query the OpenStack API to ensure that the status of the instances, volumes, and the network is consistent with the expectation of the tests. 
In our context, assertion checks serve as \emph{ground truth} about the occurrence of failures during the experiments. These checks are valuable to point out the cases in which a fault causes an error, but the system does not generate an API error (i.e., the system is unaware of the failure state) \cite{cotroneo2019bad}.

We consider an experiment as failed if at least one API call returns an API error or if there is at least one assertion check failure. 
Before every experiment, we clean-up any potential residual effect from the previous experiment, to be able to relate failure to the specific fault that caused it. We re-deploy the cloud management system, remove all temporary files and processes, and restore the OpenStack database to its initial state.

We developed an automated tool to scan the source code of Nova, Cinder, and Neutron to find all the injectable API calls, and to introduce faults by mutating the calls \cite{cotroneo2020profipy}. For each workload, we identified the injectable locations that were covered by the workload itself, and we performed one fault injection test per covered location. 
In total, we performed \numprint{2538} fault injection experiments, and we observed failures in \numprint{1314} experiments (52\%). 
In the remaining tests, there were neither API errors nor assertion failures since the fault did not affect the behavior of the system (e.g., the corrupted state is not used in the rest of the experiment, or the error was tolerated). This is a typical phenomenon that occurs in fault injection experiments \cite{christmansson1996generation,lanzaro2014empirical}; yet, the experiments provided us a large and diverse set of failures for our analysis. 
We focus on non-tolerated faults since they are the ones of interest for the analysts. Failures point out scenarios that are not yet handled by the cloud system, and that require additional fault tolerance mechanisms. The purpose of the proposed approach is to ease the identification of these failure modes.

Table~\ref{tab:dictionary_wl} shows, for each fault-injection campaign, the total number of experiments that experienced a failure, the number of events that can be observed in the distributed system during the execution of the workload under fault-free conditions, and the average length of the fault-free sequences in terms of the number of events in the recorded trace.

The number of unique events (i.e., different types of operations performed by the system) and the number of events (i.e., total operations of the system) per trace reflects the extent and diversity of the workloads. 
DEPL is the most stressful one in both regards, followed by NET and by STO.
Moreover, the DEPL and the NET workloads are more non-deterministic than STO because the former perform a massive use of the network-related operations.
Indeed, network operations are performed by the Neutron sub-system in an asynchronous way, such as by exchanging periodic and concurrent status polls among agents deployed in the datacenter and the Neutron server. 
This behavior leads to more non-deterministic variations in the traces. 
These differences among the workloads are useful to evaluate our approach under different degrees of complexity and non-determinism.

\begin{table}[t]
\caption{Experiments in the fault injection campaigns}
\label{tab:dictionary_wl}
\centering
\begin{tabular}{
>{\centering\arraybackslash}p{1cm} >{\centering\arraybackslash}p{1.5cm} >{\centering\arraybackslash}p{2.1cm} >{\centering\arraybackslash}p{1.9cm}}

\toprule
\textbf{Workload} & \textbf{Num. of exps. with failure} & \textbf{Num. of unique events in the fault-free traces} & \textbf{Avg. num. of events per fault-free trace}\\ 
\midrule
DEPL    & 537   & 64   & 285 \\ 
NET     & 262   & 40    & 252 \\ 
STO     & 515   & 41    & 109 \\ 
\bottomrule
\end{tabular}
\end{table}

In our implementation, we adopt the \emph{Zipkin} distributed tracing system \cite{zipkin}, due to its maturity, high performance, and support for several programming languages. 
The instrumented APIs send data via HTTP to a collector, which stores trace data. 
The collected events are ordered with respect to the timestamp given by the Zipkin collector.
To collect the events, we instrumented the following communication points:

\begin{itemize}
    
    \item The \emph{OSLO Messaging library}, which uses a message queue library to exchange messages with an intermediary queuing server (RabbitMQ) through RPCs. These messages are used for communication among OpenStack sub-systems.
    
    \item The \emph{RESTful API libraries} of each OpenStack sub-system, i.e.,  \emph{novaclient} for Nova (implements the OpenStack Compute API \cite{computeAPI}), \emph{neutronclient} for Neutron (implements the OpenStack Network API \cite{networkAPI}), and \emph{cinderclient} for Cinder (implements the OpenStack Block Storage API \cite{storageAPI}). These interfaces are used for communication between OpenStack and its clients (e.g., IaaS customers).
    
\end{itemize}

Zipkin puts a negligible overhead in terms of run-time execution, as it adopts an asynchronous collection mechanism to avoid impacting critical execution paths. Moreover, we only needed to instrument $5$ selected lines of code (e.g., the \emph{cast} method of OSLO to broadcast messages), by adding simple annotations (the Zipkin context manager/decorator) only at the beginning of these methods (a total of 21 lines of Python code). Our instrumentation neither modified the internals of OpenStack sub-systems nor used any domain knowledge about them. 
The interested reader can find additional details about the instrumentation and the implementation in a previous work \cite{cotroneo2019failviz}.

\subsection{Evaluation Metrics}
\label{subsec:evaluation_metric}
We evaluate anomaly detection with respect to the ability to properly classify the events within a trace. 
In particular, we evaluate the \emph{false alarm rate} and the \emph{hit rate} \cite{ye2004robustness}. 
In our context, a false alarm occurs when a non-anomalous event is classified as an anomalous one, and a hit occurs when an anomalous event is correctly classified as such. 
The false-alarm rate is given by the total number of false alarms over the total number of non-anomalous events. The false-alarm rate should be as small as possible. 
The hit rate is given by the total number of hits over the total number of anomalous events. The hit rate should be as large as possible. Both metrics range between 0 and 1.

Our fault-injection experiments generated over $450$ thousands events over $2,538$ execution traces, with $109$ distinct event types (i.e., unique events). A key concern for evaluating anomaly detection is the need for a reliable ground truth about the actual label of the events (anomalous or non-anomalous). Unfortunately, manually assigning labels to such a large set of data is prone to errors and unfeasible in practice. 
Thus, we adopted an automated evaluation method and opted for conservative estimates where needed (i.e., by underestimating the accuracy of the proposed approach). 
Firstly, we build for each workload an anomaly detection model based on the LCS algorithm with 50 fault-free traces. Then, in order to define the ground-truth of the anomalies, we run the distributed system under fault-free conditions for a large number of times, generating an additional set of 500 fault-free traces, which is an order of magnitude larger than the training set of the model. Finally, we apply the LCS algorithm on these traces. 
Since these traces are fault-free, the differences pointed out by the LCS can be considered as false alarms. We record a \emph{list of false-alarm event types} by adding an event type if it caused a false alarm. In total, the list includes respectively $38$, $30$, $18$ event types for the three workloads. Instead, common events (i.e., non-anomalous) are considered as true negatives.

In our experimental evaluation, we consider an anomaly raised by a detector as a false alarm if its type belongs to the list of false-alarm event types. 
This method is very conservative since we are labeling all events of these types as false positives, even if these events could represent true anomalies for some experiments.
This approach under-estimates the ability of the VMM at identifying true anomalies since we only take into account anomalies for events that were never affected by false alarms in our initial extensive analysis. Furthermore, our classification assumes that the LCS is not affected by false negatives, thus overestimating the accuracy of the LCS approach. 

\subsection{Experimental Results}
\label{subsec:accuracy}
We aim to evaluate how the probabilistic model can prevent false alarms and, at the same time, not to discard hits. We analyzed the fault-injection experiments that experienced a failure (i.e., an API error to the clients, or a failure identified by our assertion checks). 
To provide context for the evaluation, we compare three approaches:
\begin{itemize}
    
    \item \textbf{\emph{LCS}}, the baseline approach, which just aligns and compares traces (as in existing techniques based on \emph{reference runs} \cite{hsueh1997fault,leeke2009evaluating,natella2016assessing}), without using a probabilistic model to account for non-determinism;
    \item \textbf{\emph{LCS with VMM}}, the proposed approach, which applies a Variable-order Markov Model after LCS, as discussed in \S{}~\ref{subsec:classification};
    \item \textbf{\emph{LCS with HMM}}, a different probabilistic approach, which applies a Hidden Markov Model (instead of VMM) after LCS.
    
\end{itemize}

These approaches allow us to separately evaluate the relative influence of LCS and the probabilistic models on the accuracy of anomaly detection, pointing out any improvements due to the adoption of the probabilistic model. Moreover, we compare the accuracy of the proposed approach (VMM) with respect to a traditional probabilistic model (HMM).

\begin{figure}[!t]
    \centering
    \includegraphics[width=0.85\columnwidth]{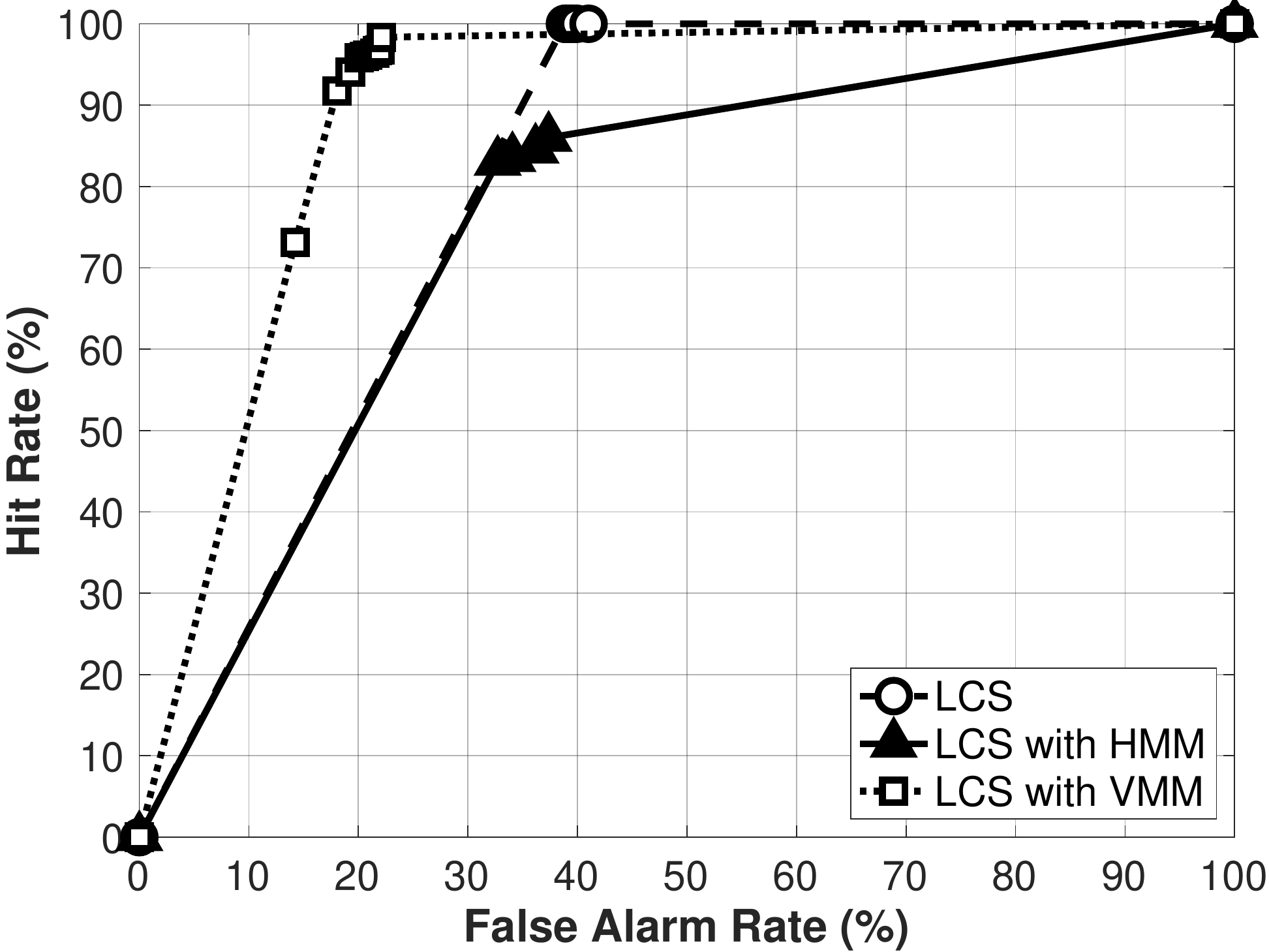}
    \caption{Approaches comparison}
    \label{fig:roc_lcs_vmm}
\end{figure}

We are interested in evaluating the accuracy of anomaly detection under different sizes of the training set (i.e., the number of the fault-free traces). 
We expect that, while increasing the number of training traces, the accuracy of the approaches improves. 
However, since false alarm and hit rates are related and often conflicting metrics, we look for trade-offs between these metrics \cite{hwang2005defending}. Thus, we use ROC curves in Figure~\ref{fig:roc_lcs_vmm} to represent both the metrics, computed over all experiments, and for different sizes of the training set between $5$ and $50$. 
Our evaluation deliberately targets the case of a limited training dataset, since it is typical for developers to have only a limited time budget to conduct test activities. In our case, an experiment takes on average 40 minutes (including the time to re-deploy OpenStack components, to revert the state of its databases and volumes, etc.), thus, 50 executions take about 33 compute hours, which we ran in parallel across several machines. If we used more training traces in our evaluation, the accuracy figures would not have been representative of what developers would achieve within a realistic amount of time.

The results show that \emph{LCS with VMM} achieves a hit rate higher than 90\%. The hit rate saturates around 98\% when the probabilistic model is trained with 20 fault-free traces, for all workloads. This size for the training set can be considered small enough for practitioners to apply the proposed approach. The proposed approach comes with a false alarm rate of around 22\%. This result means that the probabilistic model can discard many of the differences that are caused by non-deterministic behavior, even if a moderate amount of false alarms still needs to be tolerated by practitioners.

To put these results in context, we can compare them with the results for the \emph{LCS} approach. 
The \emph{LCS} achieves a perfect hit rate ($100\%$) since, with our conservative evaluation, we consider this baseline approach not affected by false negatives. The false alarm rate for \emph{LCS} is between $39$-$41\%$. The false alarm rate does not improve much by increasing the size of the training set since the LCS only identifies differences between the fault-injected trace and one \emph{selected} fault-free trace from the training set (thus, the remaining training traces do not contribute to identifying anomalies). 

The VMM is applied in pipeline after the LCS, by analyzing non-common events identified by the LCS (\S{}~\ref{subsec:classification}). Thus, the VMM can reduce the false alarm rate compared to the LCS, by classifying a ``benign'' non-common event as non-anomalous. However, the VMM can also reduce the hit rate, since it can classify a real anomaly as non-anomalous. Overall, the \emph{LCS with VMM} approach achieves a better trade-off than \emph{LCS} between a false alarm and hit rates. The loss in hit rate with respect to LCS is about 2\% since a very small number of real anomalies are discarded by the VMM. At the same time, the gain in terms of false alarm rate is quite significant, since about half of the false alarms are discarded by the VMM.

The results in Figure~\ref{fig:roc_lcs_vmm} also point out that the \emph{LCS with HMM} achieves worse performance than \emph{LCS with VMM} at identifying anomalies.
In our analysis, we carefully configured the HMM approach in order to perform a fair comparison against VMM (i.e., the one that gives the best results for HMM, in order to prevent any bias in favor of our proposed solution). To integrate HMM into our analysis, we configured the classification thresholds ($\epsilon_{SPURIOUS}$ and $\epsilon_{MISSING}$) by performing a preliminary calibration, and we selected the thresholds that achieve the lowest number of false positives without reducing the hit rate. Moreover, we varied the number of hidden states, ranging between 2 and 100. 
As in previous research that adopted HMMs, we initialized the transition and the symbol probabilities with random values \cite{qiao2002anomaly, stefanidis2016hmm}, then we used the Baum-Welch algorithm to re-estimate the parameters using the forward-backward procedure, as in the work of Batista et al. \cite{batista2010improving}. 
The ROC curve reports the results for the best configuration of the HMM approach.

Even if the HMM reduces the false alarms compared to the plain \emph{LCS} approach, the false alarm rate (about $35$\%) is still significantly higher than the \emph{LCS with VMM}. The hit rate for \emph{LCS with HMM} (about $85$\%) is also worse than the \emph{LCS with VMM}. 
We attribute this behavior to the excessive flexibility of HMMs, as they require to train a high number of parameters, which are not tuned well when using only a few tens of training fault-free traces (e.g., $50$ traces are still not enough to get a good accuracy). A similar problem would occur or be even exacerbated when using other high-dimensionality models such as neural networks \cite{nedelkoski2019anomaly}. 
Instead, even with a lower number of training traces, VMMs can achieve a better accuracy, where $20$ traces suffice to reach a good trade-off between the false alarm and hit rates.

\begin{table}[t]
\centering
\caption{Evaluation of anomaly detection, with $n=20$}
\label{tab:accuracy_table_20}

\begin{tabular}{ >{\centering\arraybackslash}p{1cm} >{\centering\arraybackslash}p{2.1cm} >{\centering\arraybackslash}p{2cm} >{\centering\arraybackslash}p{2cm}}

\toprule

\textbf{Workload}  & \textbf{Approach} &
\textbf{Avg. Hits per exp.} & \textbf{Avg. False Alarms per exp.} \\
\midrule
\multirow{3}{*}[-1pt]{DEPL} & LCS  & 14 &  92 \\
                            & LCS with HMM & 8 & 82\\
                            & LCS with VMM & 13 & 50 \\

\midrule
\multirow{3}{*}[-1pt]{NET} & LCS  & 5  & 120   \\
                           & LCS with HMM & 5 & 106 \\
                           & LCS with VMM & 5  & 58  \\

\midrule
\multirow{3}{*}[-1pt]{STO}  & LCS  & 22 & 51  \\
                            & LCS with HMM & 21 & 50\\
                            & LCS with VMM & 21 & 25   \\

\bottomrule
\end{tabular}

\end{table}

Finally, Table~\ref{tab:accuracy_table_20} shows, for each workload, the average absolute numbers of hits and false alarms per experiment, when using \numprint{20} training traces. 
It is interesting to notice that, for each workload, the number of false alarms is significantly higher than the number of hits. This difference points out that the injected faults lead to only a small number of anomalies, while the number of false alarms can be very high due to the non-determinism of distributed systems. These differences are higher for the DEPL and NET workloads that have a higher degree of non-determinism.
Moreover, the table highlights that the VMM always provides the lowest number of false alarms regardless of the workload, with a limited loss in terms of hits. 

\subsection{Sensitivity analysis}
In the previous analysis, we adopted conservative values for the VMM thresholds ($\epsilon_{SPURIOUS}=20\%$ and $\epsilon_{MISSING}=80\%$), so that the approach can filter out most of the anomalies discovered by the \emph{LCS} technique. 
Naturally, the choice of the thresholds can influence the number of false alarms and hits of the approach. 
Thus, we performed a sensitivity analysis to estimate the influence of the thresholds (\(\epsilon_{\textrm{SPURIOUS}}\) and \(\epsilon_{\textrm{MISSING}}\)) on the hit and false alarm rates. We fixed the number of training traces to $20$. 
We remark that, when the probability of a spurious event is higher than the \(\epsilon_{\textrm{SPURIOUS}}\), the event is marked as non-anomalous. Similarly, a missing event is marked as non-anomalous when its probability is lower than the \(\epsilon_{\textrm{MISSING}}\). Therefore, when \(\epsilon_{\textrm{SPURIOUS}}\) is set to $0\%$, the VMM discards all anomalies, while a \(\epsilon_{\textrm{SPURIOUS}}\) set to $100\%$ results in not discarding any anomaly. Finally, setting the \(\epsilon_{\textrm{MISSING}}\) to $0\%$ implies not to discard any anomaly, and setting the threshold to $100\%$ discards all anomalies.

\begin{figure}[!t]
    \centering
    \includegraphics[width=0.85\columnwidth]{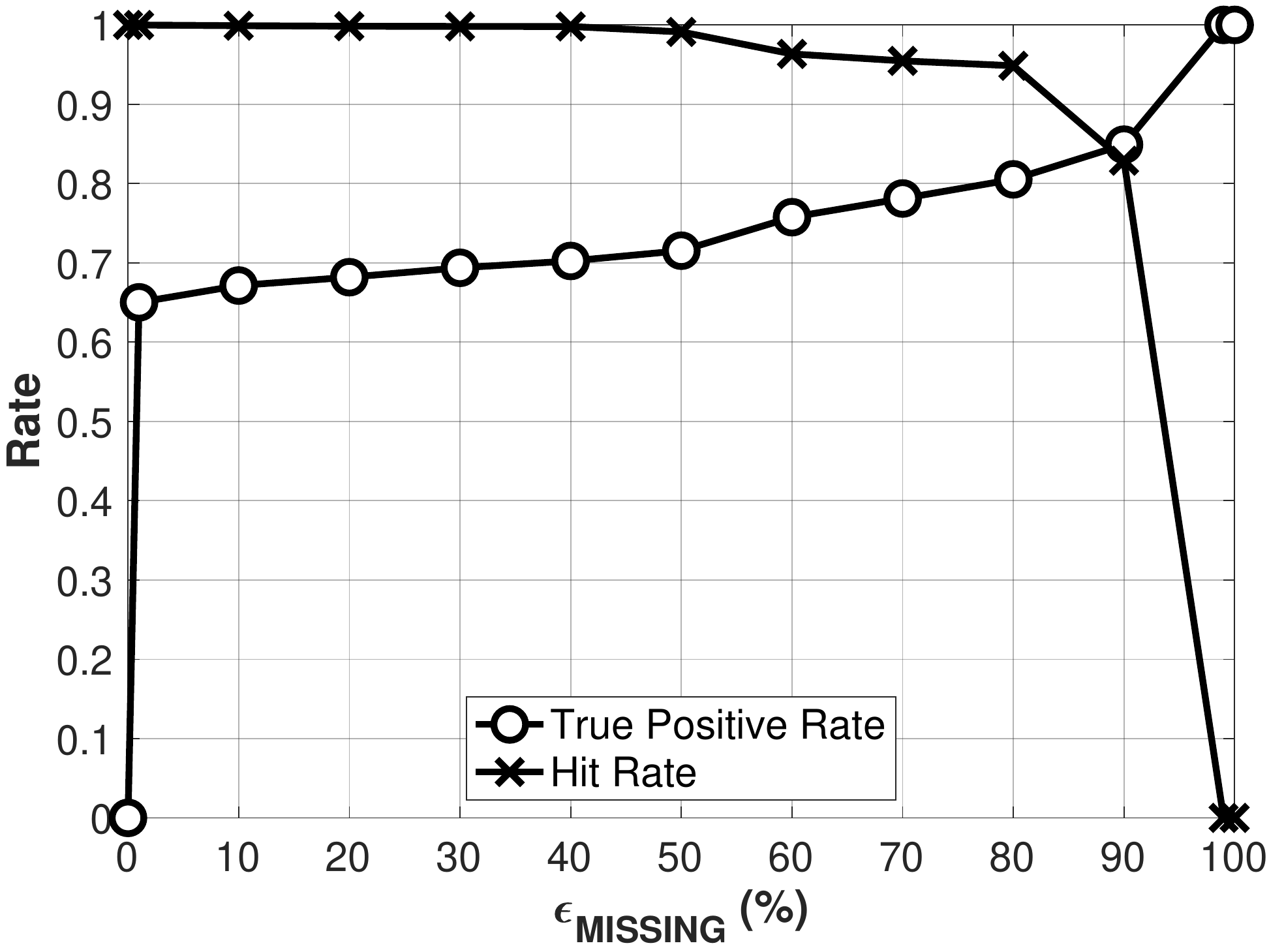}
    \caption{Sensitivity analysis for omission anomalies (\(\epsilon_{\textrm{MISSING}}\)).}
    \label{fig:ts_missing}
\end{figure}

\vspace{2pt}
\noindent
$\rhd$ \textbf{\(\epsilon_{\textrm{MISSING}}\)}.
We first analyze the accuracy of the VMM with respect to omission anomalies. Figure~\ref{fig:ts_missing} shows the rate of hits and of \emph{true positives} (i.e., the complement of false alarms, defined as $1-\textrm{false alarm rate}$, for readability), by varying the \(\epsilon_{\textrm{MISSING}}\) from $0\%$ to $100\%$.   
We can observe that the hit rate is higher than \numprint{0.99} until a value of \(\epsilon_{\textrm{MISSING}}\) equal to $50\%$. Then, the hit rate decreases slightly, until \(\epsilon_{\textrm{MISSING}}\) reaches $90$\%. Finally, the hit rate decreases rapidly to \numprint{0} at $99$\%, since even the probability of events with high-likelihood falls below the threshold. Instead, the true positive rate increases linearly after $1\%$, with significant improvement at $80\%$. Thus, $\epsilon_{MISSING} = 80\%$ is a good trade-off between hits and false alarms. The designer can fine-tune this threshold to prioritize hits over false alarms or vice versa if errors with respect to one of these metrics are not tolerated. 

\vspace{2pt}
\noindent
$\rhd$ \textbf{\(\epsilon_{\textrm{SPURIOUS}}\)}.
We performed the same analysis on \(\epsilon_{\textrm{SPURIOUS}}\), not plotted for brevity. The analysis points out that the hit rate is even less sensitive rather than \(\epsilon_{\textrm{MISSING}}\). Indeed, the hit rate only drops at $0.0$ with \(\epsilon_{\textrm{SPURIOUS}}\) equal to $0\%$, for which all anomalies are discarded. %The false alarm rate behaves similarly. 
Given that a spurious anomaly is an event that does not normally happen under fault-free conditions, the associated symbol is never encountered in the training set. The probabilistic model assigns to it a low probability since it is inversely proportional to the size of the dictionary \cite{cleary1997unbounded} and since in our experiments we collect dozens of different symbols. Thus, a conservative \(\epsilon_{\textrm{SPURIOUS}}\) (e.g., $20\%$) is a good choice since it does not impact the hits and, at the same time, discards many false alarms.

\subsection{Computational cost}
\label{subsec:performance}
\begin{figure}[!t]

    \centering
    \includegraphics[keepaspectratio=true, width=0.12\textwidth]{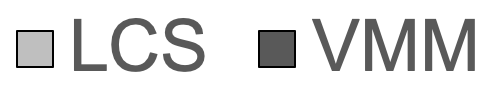}
    
    \subfloat[Number of training traces.\label{fig:time_traces}]{%
    \includegraphics[width=0.8\columnwidth]{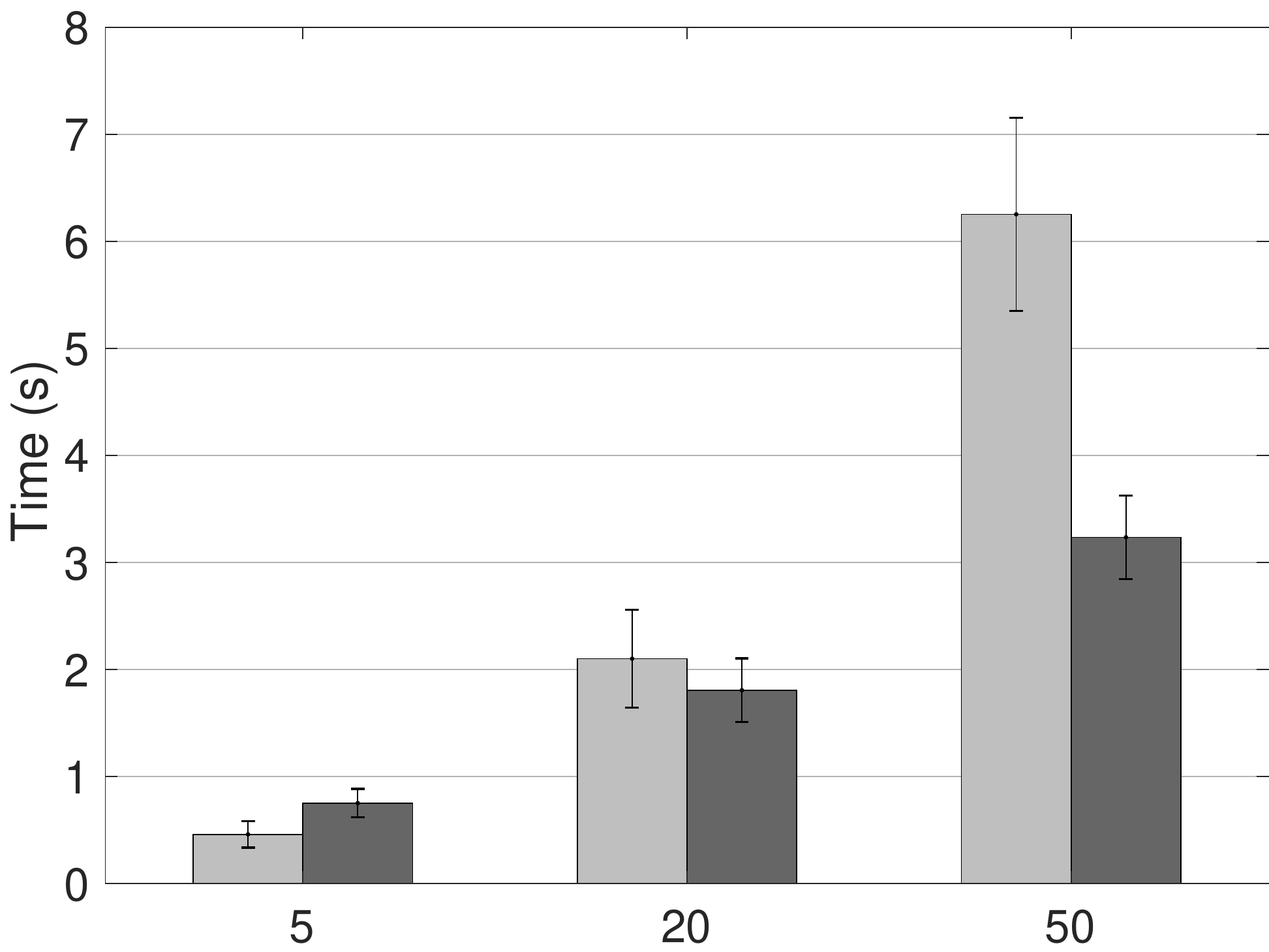}}
    
    \subfloat[Trace size.\label{fig:time_events}]{%
    \includegraphics[width=0.8\columnwidth]{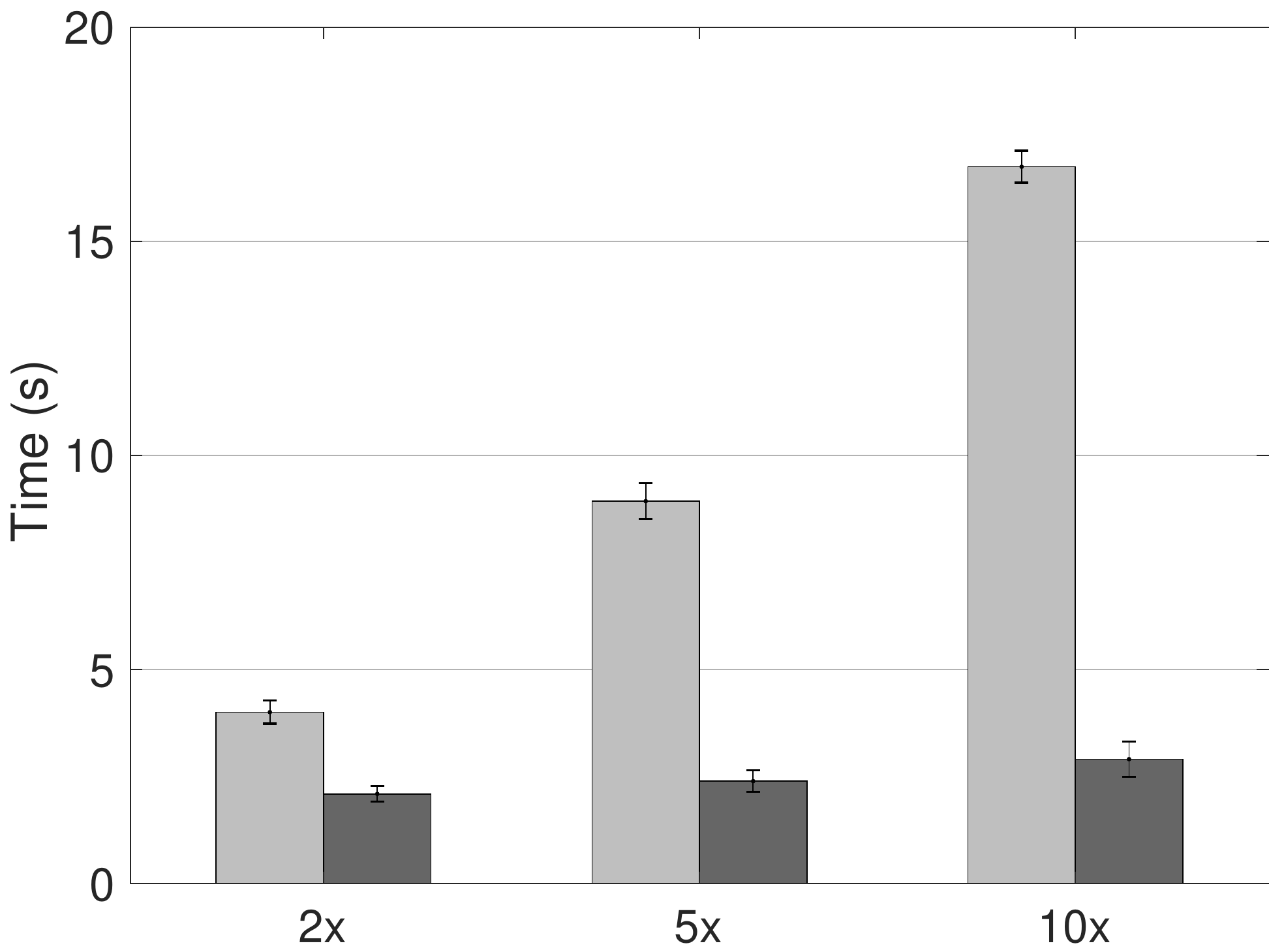}}\hfill
    \caption{Execution time for \emph{LCS with VMM}.}
    \label{fig:computational_time}
     
\end{figure}

In this section, we evaluate the computational cost and scalability of the anomaly detection algorithm. Figure~\ref{fig:computational_time} shows the time taken to analyze event traces, for increasing volumes of data, i.e., by varying the number of traces to analyze, and the number of the events per trace. 

In Figure~\ref{fig:time_traces}, we consider the average time to apply the approach on a single test trace with a fixed number of events.
The figure points out that the number of training has a higher impact on the computational time of the \emph{LCS} technique rather than the computational time of the \emph{VMM} technique. 
Indeed, the most of the time for analysis is incurred because of the search for the \emph{selected fault-free trace}, i.e., the training trace most similar to the one under analysis (see also \S{}~\ref{subsec:trace_comparison} and \figurename{}~\ref{fig:approach_flowchart}). Once the \emph{selected fault-free trace} has been found, the VMM algorithm can be executed very quickly, taking about $3s$ with $50$ training traces.

Therefore, the analysis of even thousands of fault injection experiments can be performed in a reasonable amount of time. 
Since the traces can be analyzed independently from each other, they can be partitioned across several CPUs (e.g., using SMP machines): for example, in our workstation with 8 SMP cores, it takes about $40$ minutes to analyze the two thousands of traces that were produced by our fault injection experiments.

Finally, we analyze the impact on the execution time for applying the approach by varying the number of events per trace (see Figure~\ref{fig:time_events}). We consider test traces of increasing size, by replicating the same sequence of events several times ($2x$, $5x$, $10x$). 
The execution time grows linearly, as in the previous analyses. We also found that the size of the traces has a limited impact on the computational time of the \emph{VMM} technique.

\section{Evaluation of Failure Mode Clustering}
\label{sec:clustering}
 In this section, we evaluate the accuracy of the proposed approach at identifying failure modes in fault injection tests. The approach pursues this goal by \emph{clustering} the execution traces so that the human analysts can analyze the data more easily. For example, the analyst only focuses on a sample of the experiments for each cluster instead of inspecting the whole set of experiments, which would be unfeasible for large fault injection campaigns.

We evaluate both the ability to identify the number of classes in the data (i.e., how many distinct failure modes occurred in the experiments), and to assign the fault injection experiments to the classes (i.e., the failure mode to which an experiment belongs to). 
%Therefore, we applied two approaches for \textit{cluster validation} \cite{halkidi2002cluster}. 
First, we evaluate clustering according to an \emph{internal} criteria  (\S{}~\ref{subsec:clustering_internal_evaluation}), in which we assess the quality of clustering in terms of quantities that only involve the data samples. %(i.e., without an external ground truth).
Then, we assess the quality of clustering according to an \emph{external} criteria (\S{}~\ref{subsec:clustering_external_evaluation}), in which we compare the results of clustering against a reference classification of the data (i.e., an external ground truth).
The internal evaluation assesses how well the clustering algorithm can identify the number of classes, as internal criteria are also adopted by clustering algorithms to estimate the number of classes. The external evaluation assesses how well the clustering algorithm assigns the data samples to the classes, assuming that the number of classes has been given in input to the algorithm.

We perform clustering using the vector representation of executions traces based on the VMM, as in \S{}~\ref{subsec:clustering_methodology}. We adopt an unsupervised clustering algorithm, the \textit{K-Medoids} with the \textit{squared euclidean} distance measure. 
The algorithm forms clusters by minimizing the sum of the dissimilarities between objects and a reference point for their cluster. 
Differently from the classical \textit{K-Means}, which takes the mean value of the objects in a cluster as a reference point, the \textit{K-Medoids} algorithm uses a \textit{medoid}, i.e., the most centrally located object in a cluster. Thus, \textit{K-Medoids} is less sensitive to outliers than \textit{K-Means} \cite{arora2016analysis, velmurugan2010computational}.

As a reference for the evaluation, we also analyze two alternative, simpler approaches to clustering, which we refer to as \emph{LCS} and \emph{SEQ}. The \emph{LCS} performs clustering on vector representations that are similar to the approach proposed in \S{}~\ref{subsec:clustering_methodology}, but without applying the probabilistic model. Thus, evaluating \emph{LCS} gives information on the influence of the probabilistic model on clustering (e.g., due to less false anomalies, which can distort the similarity measure). 
\emph{SEQ} is a baseline approach that does not use anomaly detection: it represents each experiment with a vector of $d$ features, where $d$ is the number of symbols in the dictionary. Each feature represents the number of times that a specific symbol occurred during the execution.

We built a ground-truth for the evaluation, by performing preliminary labeling of failures. 
The problem of having a ground-truth is a quite common open problem in all the research work dealing with log analysis.
Data labeled by real system administrators represent the ideal case with the actual ground-truth, but this option requires a significant resource commitment from a company.
Therefore, we mitigated this problem by using the same data source that would be used by a system administrator for analyzing failures, e.g., by OpenStack logs, API Errors experienced by clients, assertion checks from OpenStack developers, anomalies in the traces, etc., to classify the experiments with respect to their failure modes, based on our previous experience with OpenStack \cite{cotroneo2019bad}. 
System logs are usually good indicators of system state as they contain reports of events that occur on the several interrelated components of complex systems \cite{lim2008log}. Previous works leveraged the collection of system logs as sources of data, which could be analyzed by a system to make it aware of its internal state
\cite{vaarandi2004breadth,aharon2009one,fu2009execution,makanju2011system}.
Also, to reduce the possibility of errors in manual labeling, multiple authors discussed cases of discrepancy, obtaining a consensus for the failure modes.

We found the following types of failure modes:

\begin{itemize}
    \item \textbf{Instance Failure}: The creation of the instance fails, or the instance is created but it is in an error state.
    \item \textbf{Volume Failure}: The creation of the volume and/or the attach of the volume to the instance fails, or the volume is created but it is in an error state.
    \item \textbf{Network Failure}: The creation of network resources (e.g., networks, subnets, etc.) fails.
    \item \textbf{SSH Failure}: The instance is correctly created and up, but it is not reachable.
    \item \textbf{Cleanup Failure}: The deletion of resources (previously created by the workload) fails.
    \item \textbf{No Failure}: There was no failure during the experiment. 
\end{itemize}

\tablename{}~\ref{tab:ground_truth} shows the failure modes found for each workload (i.e., 6, 4, and 4 failures mode respectively for DEPL, NET, and STO workloads) and represents our ground truth for clustering. Even if we use the same labels for the failure modes across the three workloads, each failure mode should be considered different for each workload since they involve different resources and APIs during execution (e.g., DEPL and STO have both cleanup failures, but with different behaviors). 
This classification represents our ground truth for evaluating the results of clustering.

\begin{table}[!t]
\caption{Failure Mode Classes per Workload}
\centering
%%Table 1
\begin{tabular}{cccc}
\toprule
\textbf{Failure Mode} & \textbf{DEPL} & \textbf{NET} & \textbf{STO} \\
\midrule
Instance Failure & 224 & 56 & 320\\ 
Volume Failure & 151 & - & 38\\ 
Network Failure & 52 & 30 & -\\
SSH Failure & 41 & 176 & -\\ 
Cleanup Failure & 69 & - & 157\\ 
No Failure & 539 & 299 & 386\\ 
\bottomrule
\end{tabular}
\label{tab:ground_truth}
\end{table}

\subsection{Internal Evaluation}
\label{subsec:clustering_internal_evaluation}

After performing fault-injection experiments, the human analyst first needs to get a qualitative understanding of how the system can fail under faults, i.e., to discover how many distinct failure modes the system exhibits. 
Since the analyst does not know a priori the number $K$ of failure modes, it is part of the task of our unsupervised analysis to determine this number. A common heuristic is: (i) to configure the clustering algorithm to run with a tentative value of $K$; (ii) to evaluate the ``validity'' of the clusters, in terms of low distance between samples assigned to the same cluster, and high distance between samples assigned to different clusters; and, (iii) to repeat these steps for increasing values of $K$ until the validity index reaches a ``knee'' point (i.e., the value of $K$ after which the validity index significantly drops) \cite{halkidi2002clustering}.

In this evaluation, we apply the procedure described before in the same way an analyst would do (i.e., without prior knowledge of the number of clusters). We compare the number of clusters obtained with respect to our ground truth knowledge of the failure modes (i.e., $6$ failure modes for DEPL, and $4$ failure modes for NET and STO). 
We adopt the \emph{Silhouette} index as a cluster validity technique \cite{rousseeuw1987silhouettes}, which computes the average dissimilarity between points to evaluate the cohesion of data within clusters and the separation between clusters. 
For a given cluster $\{\tau_k\}^K_{k=1}$, this method assigns to each sample $i \in \tau_k$ a measure $s_i=\nicefrac{(b_i-a_i)}{max(a_i,b_i)}$ (\emph{Silhouette width}), where $a_i$ is the average distance between the $i^{th}$ sample and all of the samples included in $\tau_k$, and $b_i$ is the minimum average distance of $i$ to all points in any other cluster. 
By averaging the Silhouette width of samples in the same cluster, and then averaging these values across clusters, we obtain a \emph{Global Silhouette value} that can be used as clustering validity index \cite{bolshakova2003cluster}.

\begin{table}[!t]
    
    \centering
    \caption{Number of clusters using the \emph{Silhouette} index, with different clustering approaches.}
    \label{tab:silhouette_table}

    \begin{tabular}{ >{\centering\arraybackslash}p{1cm}
    >{\centering\arraybackslash}p{1cm}
    >{\centering\arraybackslash}p{1cm} >{\centering\arraybackslash}p{1cm} >{\centering\arraybackslash}p{1.5cm}}

    \toprule
    \textbf{Workload} &
    \textbf{Actual clusters} & \textbf{SEQ} & \textbf{LCS} & \textbf{LCS with VMM}\\
    \midrule
    DEPL & \textbf{6} & 2 & 6 & 6\\
    \midrule
    NET & \textbf{4} & 5 & 3 & 5\\
    \midrule
    STO & \textbf{4} & 4 & 3 & 4\\
    \bottomrule
    \end{tabular}
\end{table}

We configure the clustering algorithm with tentative values for the number of $K$ clusters, with values between $K=2$ and $K=20$. 
Table~\ref{tab:silhouette_table} shows the number of clusters suggested by the \emph{Silhouette} index, for the three vector representations and the three workloads. 
In the case of clustering based on \emph{VMM}, the ``knee'' point matches, or is very close, to the number of clusters in our ground truth, for all of the three workloads. The other two clustering approaches (i.e., \emph{LCS} and \emph{SEQ}) are only accurate for some workloads but do not perform well for other ones. For example, in the case of the DEPL workload, the knee point at $K=2$ for \emph{SEQ} is much lower than the actual number of clusters $K=6$ in our ground truth. For the NET and STO workloads, the validity index for \emph{LCS} drops at $K=3$ clusters, but clustering should find at least $K=4$ clusters according to the ground truth. Overall, the vector representation with VMM leads to a more reliable indication of the number of clusters.

\subsection{External Evaluation}
\label{subsec:clustering_external_evaluation}

The external evaluation assesses clustering algorithms as in a classification problem, by comparing the clusters with respect to the failure modes in our ground truth (\tablename{}~\ref{tab:ground_truth}). We compare, for each element in the dataset, the cluster assigned to the element with the actual class of the element, according to the ground truth. We adopt the following rule for the comparison \cite{modha2003feature}: for every cluster generated by the algorithm, we identify the ground-truth class with the largest overlap and assign every element in the cluster to the ground-truth class. In the case of a poor clustering algorithm, multiple clusters may be assigned to the same ground-truth class, but it never assigns the same cluster to multiple ground-truth classes.

In quantitative terms, let $C$ be the number of ground-truth classes $\{\omega_c\}^C_{c=1}$. 
The \emph{purity} of a cluster is defined as the fraction of elements in the cluster that matches the ground-truth class  \cite{xiong2009k}. Assuming $K$ clusters, for each cluster $\{\tau_k\}^K_{k=1}$ we define $P_k = \nicefrac{1}{n_k} \, \cdot max(n^{c}_k)$, where $n_k$ is the size of the cluster $\tau_k$, and $n^{c}_k$ is the number of elements in the cluster $\tau_k$ that belong to the class with label $w_c$. 
The overall \emph{purity} achieved by a clustering algorithm is the weighted sum of purities across classes, given by $P = \sum\nolimits_{k=1}^K \nicefrac{n_k}{n} \, \cdot P_k$. 
The larger the value of purity, the better the clustering quality.

\begin{table}[!t]
\centering
\caption{Purity of clusters, with different techniques.}
\label{tab:purity_table}

\begin{tabular}{ >{\centering\arraybackslash}p{1cm} >{\centering\arraybackslash}p{1.5cm} >{\centering\arraybackslash}p{1.5cm} >{\centering\arraybackslash}p{1.5cm}}

\toprule

\textbf{Workload} &
\textbf{SEQ} & \textbf{LCS} & \textbf{LCS with VMM}\\
\midrule
DEPL  & 0.74 & 0.91 & 0.94\\
\midrule
NET  & 0.85 & 0.81 & 0.86\\
\midrule
STO & 0.82 & 0.86 & 0.90\\

\bottomrule
\end{tabular}
\end{table}

We compute for each workload the purity obtained by the three clustering techniques. 
Table~\ref{tab:purity_table} shows the results. We perform $50$ repetitions and compute the average value of purity across repetitions. We omit the standard deviation since it is negligible (lower than $1\mathrm{e}{-03}$). 
The results suggest that, for all workloads, the \emph{LCS with VMM} always provides the highest purity value.  
Moreover, we can notice that the VMM leads to an increase in the value of purity ranging between $3\%$ and $5\%$ when compared to the basic \emph{LCS} approach. 
The \emph{SEQ} technique leads to worse results, especially in the case of a very stressful workload such as DEPL, where the sequence of events is longer and with more types of events. 
We performed the statistical hypothesis test (\emph{Student's t-test}) to verify that differences are statistically significant: this is indeed the case, as the test rejects the null hypothesis at the $1\%$ significance level.
%with high confidence (p-value lower than 0.01).}
Thus, the proposed probabilistic model can enhance the accuracy of failure mode clustering.

\section{Related Work}
\label{sec:related}
%### FAULT INJECTION IN DISTRIBUTED SYSTEMS ###

\noindent
\textbf{Uncertainty in fault injection experiments.} 
Uncertainty is a pervasive aspect in fault injection experimentation, as the behavior of complex systems depends on many factors that difficult or impossible to control. Traditionally, these factors have been addressed using statistical techniques: for example, previous studies on hardware fault injection have been sampling the space of fault injections (i.e., CPU instructions and data words to be injected with \emph{bit-flips}), and applying statistical modeling for the probability of failures (e.g., to obtain confidence intervals) \cite{arlat2011collecting, skarin2010goofi, palazzi2019tale}. This approach has been generalized in the AMBER project \cite{wolter2012resilience}, which adopted data mining techniques to analyze large sets of experiments, to identify which factors (e.g., the type of injected faults, the workload, the configuration of the target system, etc.) has the highest impact on performance and availability. 

These concerns also apply to distributed systems, where non-deterministic behavior introduces additional uncertainty in experimental results. To address them, Bondavalli et al. \cite{bondavalli2007foundations, bondavalli2010new} applied the principles of \emph{measurement theory} to assess the quality of measurements in terms of uncertainty, repeatability, resolution, and intrusiveness. 
The Loki tool \cite{chandra2004global} addressed the problem of injecting faults in controlled global states of distributed systems since it is difficult due to the lack of a global clock and communication delays (e.g., between a central controller and a local injector). Thus, Loki performs a post-experiment analysis of event traces collected from nodes, using an off-line clock synchronization algorithm, to identify whether injections actually hit the desired state, and repeats the experiments only when needed. Overall, these studies assume that failures can be automatically and accurately identified, and our work complements them by providing techniques for identifying the failure modes.

\vspace{3pt}
\noindent
\textbf{Property checking.} 
The existing fault injection tools detect the occurrence of failures by looking for specific events, such as service errors returned by the distributed system to its clients (e.g., API errors); performance degradation and bottlenecks; high-severity error messages in the logs of the system; and assertion failures introduced by developers inside the software. 
\emph{Destini} \cite{gunawi2011fate} uses a declarative relational logic language (Datalog) to allow developers to customize \emph{test specifications} (i.e., fault-tolerance properties that need to be fulfilled in the presence of faults), and for checking that the system complies with them. 
These specifications are expressed in terms of events (e.g., failures and protocol events), and relations over them representing expectations and facts (e.g., data blocks or packets that are expected in a given state, which are compared with the ones that are actually observed during the test). 
Similarly, \emph{P\#} \cite{deligiannis2016uncovering} identifies failures using liveness specifications (e.g., lack of progress, such as the inability to restore a failed node) and safety specifications (validity assertions on the local and global states of the system), written with a domain-specific language in terms of communicating state machines with asynchronous events. 
However, these solutions require domain expertise and human effort to be applicable. In this work, we investigate techniques to automate the identification of failure modes without supervision, to ease the adoption of fault injection by practitioners.

%### TRACE ANALYSIS OF DISTRIBUTED SYSTEMS ###

\vspace{3pt}
\noindent
\textbf{Execution trace analysis in distributed systems.} 
Research studies on debugging distributed systems lead to a variety of tracing-based techniques. 
For example, Magpie \cite{barham2003magpie}, Pinpoint \cite{chen2004path}, and Aguilera et al. \cite{aguilera2003performance} identify causal paths in the distributed system, by tracing and correlating call requests and responses, and events at the OS-level and the application-server level. 
These approaches were still too difficult to apply in practice, as they either focused only on synchronous (RPC-style) interactions between components and neglected asynchronous and concurrent ones; or, they required intrusive instrumentation of the entire software stack down to the OS. 
Pensieve \cite{zhang2017pensieve} is an approach similar way to delta debugging, to reconstruct the intermediate path backward from the failure to the user inputs and events that cause the failure, by combining static analysis and iterative re-executions of the system. 
Friday \cite{geels2007friday} is a distributed debugger that allows developers to replay a failed execution of a distributed system, and to inspect the execution through breakpoints, watchpoints, single-stepping, etc., at the global-state level. 
ShizViz \cite{beschastnikh2016debugging} is an interactive tool for visualizing execution traces of distributed systems, which allows developers to intuitively explore the traces and to perform searches; moreover, the tool provides support for comparing distributed executions with a pairwise comparison, even if without probabilistic techniques to filter-out benign variations due to non-determinism.

The approach proposed in this paper is unique in the design space of distributed system analysis, as it is the first one tailored for the analysis of fault injection experiments. 
The approach identifies failures without relying neither on programmer-written specifications nor on intrusive instrumentation, and it is applicable even in the presence of asynchronous and rare interactions. Our approach can be easily deployed and integrated into interactive tools for debugging and visualization, to provide more robust trace comparison and analysis abilities.

\section{Conclusion}
\label{sec:conclusion}
In this paper, we presented a novel approach for discovering failure modes in distributed systems, by combining fault injection, distributed tracing, and unsupervised learning algorithms. 
By adopting a probabilistic model (VMM), our approach can identify anomalies in noisy execution traces, by significantly reducing the false alarms without discarding true anomalies. To further help the human analyst at analyzing failures, we presented a novel technique that clusters fault injection experiments according to classes of failure modes. The results showed that clustering can achieve high accuracy under different conditions. 
Future works will include an analysis of how the training of a model in a specific execution environment affects the discovering of failure modes from fault injection data obtained from different execution environments.

\ifCLASSOPTIONcaptionsoff
  \newpage
\fi

\bibliographystyle{IEEEtran}
\bibliography{bibliography}

\vspace{-1cm}
\begin{IEEEbiographynophoto}{Domenico Cotroneo} (Ph.D.) is an associate professor at the University of Naples Federico II, Italy. His research interests include software fault injection, dependability assessment, and field-based measurement techniques.
\end{IEEEbiographynophoto}
\vspace{-1cm}
\begin{IEEEbiographynophoto}{Luigi De Simone} (Ph.D.) is a postdoctoral researcher at the University of Naples Federico II, Italy. His research interests include dependability benchmarking, fault injection testing, virtualization reliability and its application on safety-critical systems.
\end{IEEEbiographynophoto}
\vspace{-1cm}
\begin{IEEEbiographynophoto}{Pietro Liguori} is a Ph.D. student at the University of Naples Federico II, Italy. His research activity includes anomaly detection, failure mode analysis, and software fault injection in cloud computing infrastructures. His research interests also focus on neural machine translation to automatically generate software exploits.
\end{IEEEbiographynophoto}
\vspace{-1cm}
\begin{IEEEbiographynophoto}{Roberto Natella} (Ph.D.) is an assistant professor at the University of Naples Federico II, Italy. His research interests include dependability benchmarking, software fault injection, software aging and rejuvenation, and their application in OS and virtualization technologies.
\end{IEEEbiographynophoto}

\end{document}